# A Survey on Segment Routing with Emphasis on Use Cases in Large Provider Networks

Aniruddha Kushwaha, Sidharth Sharma and Ashwin Gumaste
Indian Institute of Technology Bombay, India

***Abstract*—Segment routing is heralded as an important technology innovation in large provider networks. In the domain of large telecom service providers, segment routing has the potential to be in the same league as MPLS and IPv6 as well as pave a way towards the successful implementation of SDNs. In this regard this paper is a survey on the existing segment routing work in the community. We begin by describing how segment routing works, inclusive of the various building blocks. The paper describes the segment routing architecture in terms of its positioning vis-à-vis existing networking technologies, as well as the data-plane and the control plane. Label encoding techniques that are central towards implementing segment routing are then discussed. In order to make segment routing relevant to the current domain of providers, we postulate various use cases, their working and issues of interoperability. The paper also reviews the current set of segment routing implementation cases in industry. Thereafter we position segment routing among the various provider manifestations.**

***Index Terms*—segment routing.**

## I. Introduction

Internet traffic is doubling almost every other year due to introduction of new services and applications such as Video-on-Demand, Voice-over-IP, and over-the top (OTT) services like YouTube, Instagram, Facebook, etc. It is becoming important to route, provision, transport this enormous traffic efficiently in large service provider networks by satisfying various service level agreements (SLAs) in terms of latency, packet loss, jitter, bandwidth, etc. and at low price-points. Traditionally, service provider networks use IP-based forwarding and a distributed control plane as well as constrained shortest-path forwarding (CSPF) as methods for implementing intra-domain routing. Interior gateway protocol (IGP) and border gateway protocols (BGP) are used to distribute the shortest path and policy-based topology information respectively, to the nodes in the network. Traffic engineering is a way by which congestion in links can be avoided by spreading traffic across various links.

One of the key enabling technologies for traffic engineering is MPLS [1]. MPLS along with Carrier Ethernet [2] is typically being deployed for data transport in large provider networks. The problem with MPLS is that it is per se not carrier-class and even variants of MPLS such as MPLS-TP (which is actually Carrier Ethernet) [3] are not convenient in providing a scalable, distributed packet-core at low price-points. The key question that manifests due to MPLS deployment is to create a scalable packet core that is carrier-class, traffic-engineered with a centralized control plane. The latter being the key towards inculcation of future SDN concepts in provider networks. With the control plane moving towards a centralized manifestation, and groups of nodes getting intelligence from a controller/network management system, a key technology that can be used for achieving carrier-class transport as well as traffic engineering is source-routing [4]. Using the building blocks of the MPLS data-plane, a centralized control and source routing, the question then gears towards creation of a unified packet core. To this end, segment routing was proposed as an add-on technology that seamlessly integrates with MPLS, IPv6, Carrier Ethernet [3] as well as with other packet-optical manifestations [5].

Segment routing or SR, is a source-routing concept that facilitates the creation of segments that point towards sub-paths along a larger path. Conjoining multiple segments at the source or across a path, facilitates an end-to-end implementation. The advantage of SR is that it can continue to leverage an existing distributed system, while interacting with a centralized management framework that works well with future SDN implementations. In case of SDN based segment routing networks a centralized controller can compute segment identifiers and sub-path information, while the data-plane implementation is left to the different forwarding planes that are used. This approach can potentially inculcate programmability and eventually lead to scalability. Segment routing is a betterment over its predecessor, MPLS, especially in the provider domain, where carrier-class performance is key towards revenue growth and provisioning OTT services. It can be said that segment routing is a promising futuristic technology for data-transport in the WAN, especially as SDN graduates from the data-center towards the WAN. In this context, SR manifestations as products, technologies and ideas are key to study from the perspective of future deployments.

### A. Building blocks and motivation for segment routing

*MPLS:* The first major departure from IP, was the MPLS scheme. The MPLS architecture [1] came in the late 1990s as an alternative to traditional IP routing. The first goal of MPLS was to facilitate forwarding based on labels, which are network specific as opposed to IP-addresses, which were largely global. This implied MPLS would be faster, and forwarding tables would be easier to build and operate. MPLS was designed to work between layer-2 and layer-3 and therefore is often described as a layer 2.5 protocol. Compared to IP-routing, where an IP lookup is used to find the correct path at each of the individual router, MPLS works on a label-based switching scheme, where the label value (a 20-bit value) is relevant to the domain of the MPLS network. A label is pushed into the packet header in between the Layer 2 and Layer 3 headers at the ingress router. A label is created for a path known as a tunnel between an ingress and egress node. All routers along the path use this label to identify the tunnel for forwarding. MPLS introduced the creation of a traffic engineered (TE) path known through the MPLS-TE paradigm [6] and this facilitates end-to-end (E2E) explicit routing based on the establishment of a dedicated tunnel between ingress and egress nodes. The objective of TE is to manage traffic such that the network capacity is efficiently utilized. Therefore, E2E explicit routing may consider any path of choice based on user-defined parameters, and this path is not necessarily the shortest path. Although traffic engineering led to avoidance of congestion along the shortest path by providing alternate routes, there were several challenges for operators in order to implement TE. These challenges are: control plane overhead and scalability related problems [7]; requirement of a dedicated signaling protocol for tunnel management i.e. the resource reservation protocol-traffic engineering (RSVP-TE) [8], label distribution protocol (LDP) [9] and; creation of flow states at the intermediate routers. When the number of tunnels in a network became large, there is a requirement of a large forwarding table that resulted in increased the processing time for packet forwarding.

*IPv6:* Due to the massive explosion of users and devices the IPv4 address range has nearly exhausted. To overcome the IPv4 address exhaustion, the IETF developed early on RFC 2460 [10] and proposed IPv6. IPv6 uses 128-bits for addressing, thereby allowing $2^{128}$ addresses. The IPv6 header was designed to offer more flexibility compared to IPv4 such as cater to security, routing, fragmentations, authentication etc. These features were incorporated by allowing extensions to the IPv6 header [10]. For example, the routing extension of IPv6 allows to include the intermediate nodes to be visited by a packet in the header. Therefore, this routing header extension makes IPv6 suitable for source routing.

*Source-Routing:* Source-routing was initially suggested by Farber and Vittal in [11] and discussed by Sunshine in [4] as early as 1970s. Source-routing provides for complete path information at the source itself using which a packet can traverse a path towards a destination. This information is encapsulated in a packet at the source node (router) itself. However, the main challenge at the time of conceptualization and implementation of source-routing is that of path selection by the source as this required global and instantaneous knowledge of the entire network. Source routing hence requires significant centralized processing, as compared to other protocols such as OSPF etc., which require distributed processing. As the network evolved, new protocols such as IGP, BGP and their extensions came into existence which were used by the routers to distribute network state information with each other. Since, this was cheaper and easier to implement, therefore, source routing was not much adopted beyond select unconventional or specific network applications.

Source-routing has recently resurfaced thanks to many vendors proposing a centralized management system. Such centralized control/management systems are being used to manage large providers today. A provider domain with 100s of core nodes or 1000s of metro and core nodes is often managed through a centralized control entity. Such an entity is particularly useful for traffic engineering. This centralized control entity is also a benefactor towards the evolution of SDNs. Segment routing is hence a way of using source routing along with technological advances of traffic engineering at the control level, MPLS, IPv6 and Carrier Ethernet at the data-plane level.

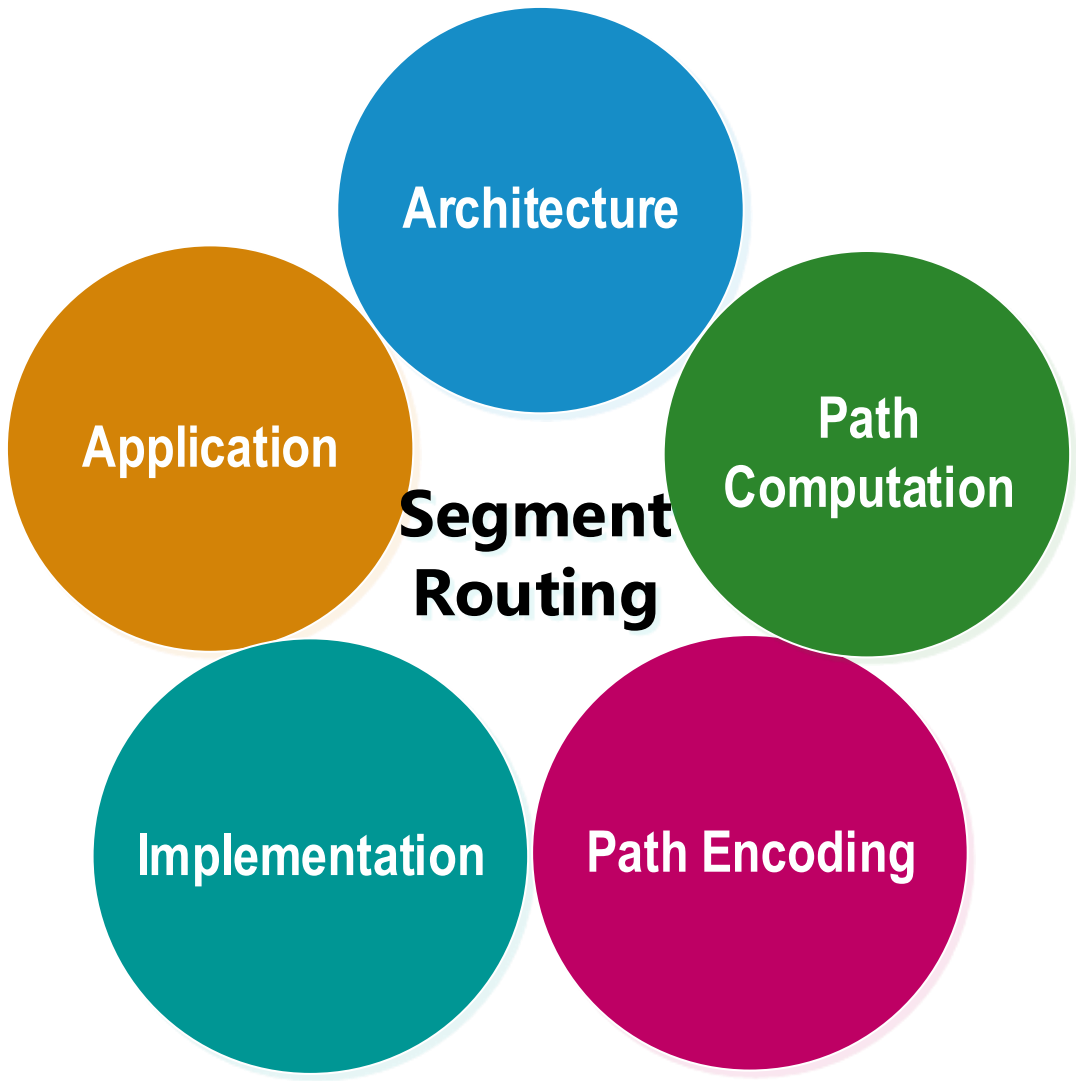

Fig. 1. Segment routing research areas.

*What is Segment Routing:* Segment routing (SR) is a recently proposed technique to provide an effective way of doing traffic engineering. SR is a source-routing based tunneling technique that allows a source (ingress) router to steer a packet through a network by embedding path information in a packet as a list of segments. A segment can be an intermediate node or a link connecting two nodes. A segment signifies the passage of packets/traffic through designated link/node in the network. Existing/ongoing research in segment routing can be broadly classified in five categories as shown in Fig. 1. These five categories are: architecture, path computation, path encoding, implementation and application, which are inter-linked to each other. In this paper, we discuss all five areas and compare different approaches within the category.

In its current manifestation, the SR architecture supports both distributed and centralized control to provide an efficient solution. In case of distributed control, a node uses the distributed information for the computation of segments. A node efficiently re-optimizes the segments based on events such as fault, congestion, re-routing, etc. The full potential of SR can be exploited with centralized control by performing global optimization and choosing the best path for a packet and forwarding the packet based on a global view of the network. The centralized control element can be a Path Computation Element (PCE) or Software Define Network (SDN) controller (for provider networks) or it can be a manual operator's network management system (NMS) for small networks. Since SR uses extensions of IGP routing protocols therefore, it does not require an additional signaling protocol such as LDP or RSVP-TE. This results in reduction in control plane complexity (compared to MPLS). It has been shown in [12] that SR provides TE performance comparable to a MPLS network.

Segment routing (SR) facilitates the end-to-end (E2E) source-based routing where the complete path to be followed by a packet is known and encoded in the packet at the source (ingress) node. The SR framework [13], [14] provides efficient traffic engineering capability that can be exploited to reduce the traffic over highly utilized links in a network and also to optimize network utilization. Unlike MPLS-TE [6], segment routing does not add any complexity in terms of a signaling protocol that is required to carry network information. The absence of an explicit signaling protocol makes the control plane of the SR framework simpler and easy to implement. In one embodiment, segment routing ensures that nodes get a global view of a network, by using an existing IGP protocol. A node thereafter computes an end-to-end (E2E) SR path as a list of segments. A segment represents a sub-path of an E2E path. A computed list of segments is thereafter embedded in a packet in the form of an SR header by the ingress node. This makes the system conform to source-routing. Subsequently, intermediate nodes along a path extract the active segment information from the packet and steer the packet along the chosen path towards the destination. This results in lowering of data-plane complexity.

Segment routing allows to enforce end-to-end policies in a network by avoiding complex protocol conversions. Though SR can be implemented without SDN, a centralized control plane such as the one juxtaposed by SDN is desirable to utilize the full potential of SR. Segment routing with SDN provides an opportunity to an operator for orchestrating services across multiple domains, where each domain is operating a different routing protocol. In segment routing, forwarding rules are stored at the ingress routers, therefore, which improves the time for network convergence. In this sense, SR is a much simpler, scalable and flexible protocol than any other SDN protocol (such as say OpenFlow [15]) and SR features can be easily implemented in software without any changes to existing hardware. This makes SR favorable for SDN implementations in provider networks, with an SR application that can now be potentially run through the North Bound of an SDN domain. In fact, even new SDN techniques like P4 have segment routing implementations [16] associated with them. It is expected that the northbound interface of the SDN controller can provide ample opportunity for providers to utilize segment routing for new applications.

In this paper, we present a comprehensive literature survey on segment routing with main focus on the segment routing use-cases. We begin with the segment routing architecture and discuss the components of the SR framework in section II, which also includes discussion from various RFCs and IETF drafts. In section III, we discuss and compare the different path computation and SR label encoding schemes. Section IV presents the use-cases of SR. We discuss various SR implementations in the section V. In section VI, we explain the use of SR in the commercial equipment. In section VII, we discuss the positioning of SR in different networking scenarios and the discussion in section VIII concludes the paper.

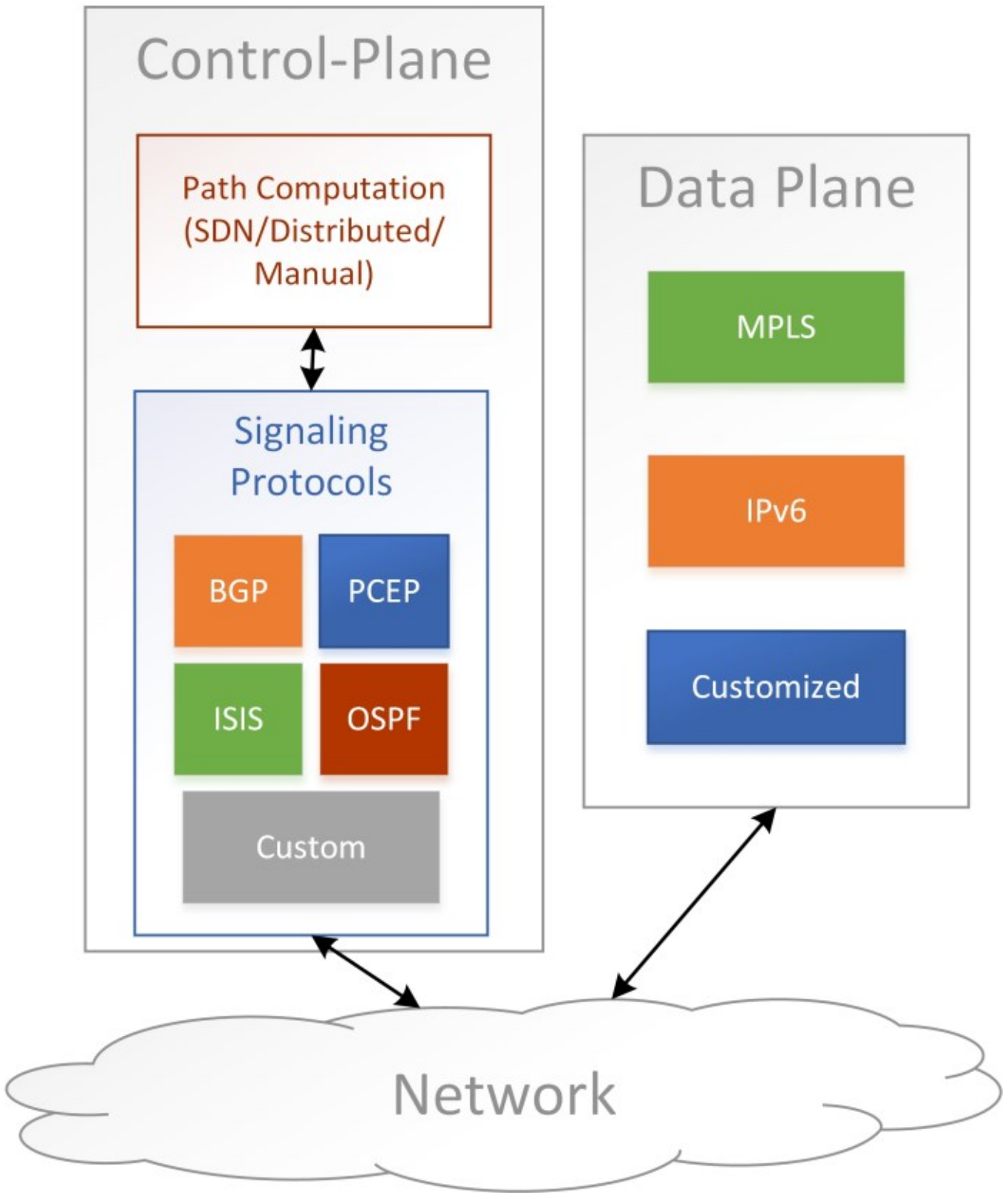

Fig. 2. Segment routing framework.

## II. Segment Routing Architecture

In this section, we describe the segment routing framework as shown in Fig. 2. This framework consists of two components: a data-plane and a control-plane.

### *A. Segment routing data-plane*

The data-plane of the segment routing framework defines: a) the labeling of the segments; b) the encoding of the segments in a packet; c) the processing of the segments by a segment routing enabled node and; d) SR data-plane implementation. We shall now describe these in detail.

#### *1) Segment labeling*

The source node determines the path as a sequence of segments. This path may contain single or multiple segments. A segment represents a shortest path to the intermediate/destination node. Each segment is associated with a label known as a segment identifier (SID). There are three types of SIDs:

- Node-SID: This SID is globally unique. A Node-SID is assigned to each node of the network by a centralized SDN controller or can be manually assigned by a network operator. A segment with a particular Node-SID at a node implies that a packet can be forwarded along the shortest path to the node specified by that particular Node-SID. In case of the MPLS data-plane, this SID corresponds to an MPLS label.
- Adjacency-SID: This type of SID is of local significance and a node assigns an Adj-SID to each of its interfaces. Adj-SID is generally followed by a Node-SID, where the Node-SID is used for forwarding a packet to a specified node and then the Adj-SID is used at the specified node to identify the interface to further forward the packet.

- Service-SID: This type of SID is also of local significance. A service SID is assigned to each service type provisioned at the node i.e. point-to-point, point-to-multipoint, VPN, firewall, deep-packet inspection etc. Similar to Adj-SID, service-SID is also followed by the Node-SID, where the Node-SID allows the packet to reach a specified node and then the Service-SID is used to deliver the packet to the particular service running over the node.

*2) Segment encoding and SR working*

The list of segments for a path towards a destination node are encoded as the SR header in the packet by an ingress node. The SR header consists of: i) list of segments specifying the path to be taken towards the destination and, ii) a pointer to indicate an active segment. Every node in the network has a forwarding table, in which the shortest-path information to all other nodes is stored. At the ingress node in addition to shortest path information, respective instructions to modify the packets is also stored. An SR node supports three types of instructions: *Push*, *Next* and *Continue*.

The *Push* instruction is used to add a segment ahead of the present header segment, resulting in the added segment as the active segment.

The *Next* instruction is used to mark the next segment as the active segment.

The *Continue* instruction is used to make a forwarding decision based on an active segment.

*3) Segment processing and SR working*

Upon reception of a packet with an SR header, the node examines the active SID. In case the SID is of the node itself, then, the *Next* instruction is executed to point towards the next segment. The next SID in the list can be a Node-SID or an Adj-SID. For a Node-SID, the node performs a lookup in the forwarding table to identify the shortest path for the node specified by the SID. For Adj-SID, the node forwards the packet to an associated interface and marks the next segment as an active segment by executing the next instruction. Once an egress node is reached, the SR header is stripped off from the packet.

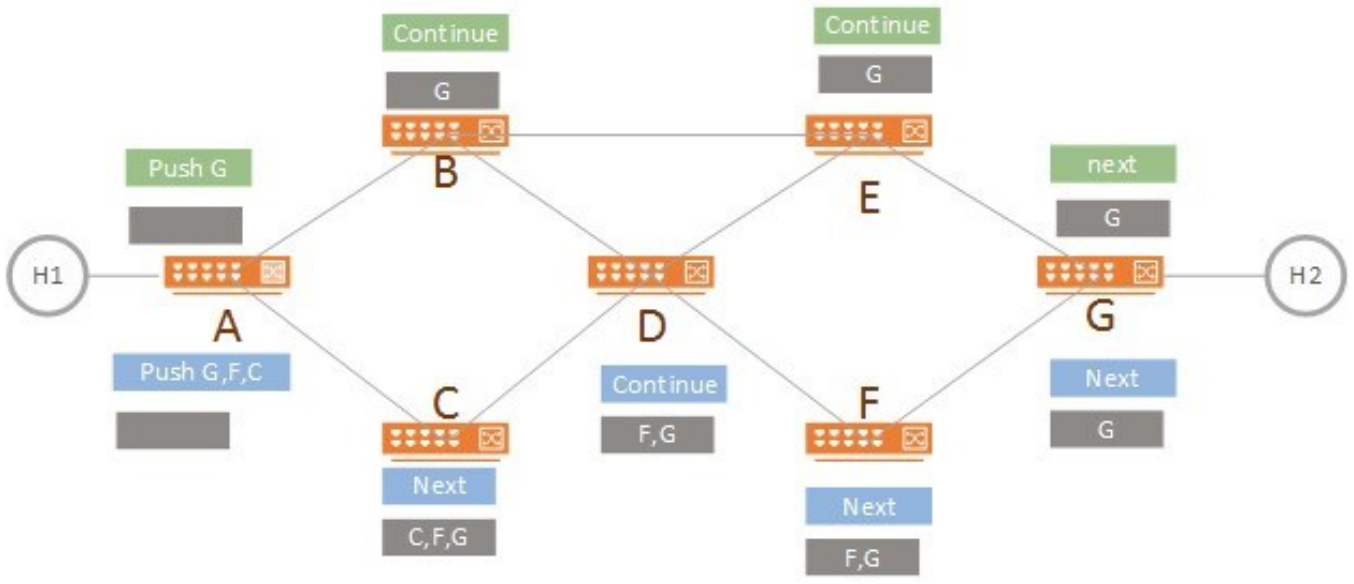

Fig. 3. Segment routing illustrations.

Shown in Fig. 3 is an SR network where two hosts H1 and H2 are connected to node *A* and *G* respectively. There are multiple paths available between nodes *A* and *G* but initially only the shortest path is considered for forwarding the packet between the nodes *A* and *G*. An incoming packet at node-*A* destined for host H2 must follow the shortest path *A-B-E-G*. Node-*A* performs a lookup in the forwarding table for obtaining an entry towards a path to H2. The lookup results in a path *B-E-G* along with a push instruction to add *G* in the list of segments. Since, there is no shortest path other than the *A-B-E-G* between nodes *A* and *G* the resultant is a single segment label that will deliver packets to *G*. Hence, node-*A* pushes the segment label *G* onto the packet.

TABLE I
COMPARISON OF THE DIFFERENT DATA-PLANES

| Parameters | MPLS [17] | IPv6 [19] | Customized SR [22] |
|---|---|---|---|
| Signaling Protocol | Centralized/ IGP extension | Centralized/ IGP extension | Centralized/ New protocol |
| SIDs | Node-SID and Adj-SID | Node-SID | Similar to Adj-SID |
| SID size | Large | Very Large | Variable |
| Header Size | Variable but medium | Variable and very large | Variable and small |
| Complexity | Medium | Medium | Low |
| Interoperability | No | Yes | No |
| Scalability | Limited | Limited | Limited |

In the event that we desire to follow the path *A-C-D-F-G*, then node *A* will push two extra segments *C* and *F* before the label *G*. Note that node-*G* has two incoming paths from node-*A*, one through node-*E* and the other through the node-*F*. Our desired path hence has *F* as an intermediate node which implies that *F* is inserted in the list of segments. Further, in order to reach node-*F* there are two available paths *A-B-D-F* and *A-C-D-F*. In order to distinguish between the two paths and follow the desired path, an additional segment is required. Therefore, node-*C* is also added in the list of segments. For every packet to follow the path *A-C-D-F-G*, a *Push* instruction with segments *G*, *F* and *C* is executed at node-*A* and *C* is marked as the active segment. There are two ways to increment the pointer: 1) the next instruction is executed at the node preceding the destination node (Penultimate hop popping) and 2) the next instruction is executed at the destination node itself. Throughout this paper, we consider and prefer the second scheme (on account of implementation flexibility), although we do note that either selection does not impact any functionality. Once the packet comes at the node-*C*, then the next instruction is executed to identify the next segment (*F*), and thereafter the shortest path *D-F* is obtained. At node-*D* a lookup for active segment node-*F* is performed. This lookup results in the shortest path towards node-*F* and a *continue* instruction is executed to keep segment *F* as the active segment. At node-*F* the *next* instruction is executed and segment-*G* is now set as the active segment. At node-*G* we repeat the execution of the *next* instruction. This is because there are no other segments remaining in the segment list. Finally, the SR header is stripped and the packet is delivered to host H2 in its native format.

*4) Segment routing data-plane implementation techniques*

We now discuss different data-plane implementation techniques that facilitate the deployment of segment routing.

The key to data-plane implementation is the manner and definition of the segment identifier.

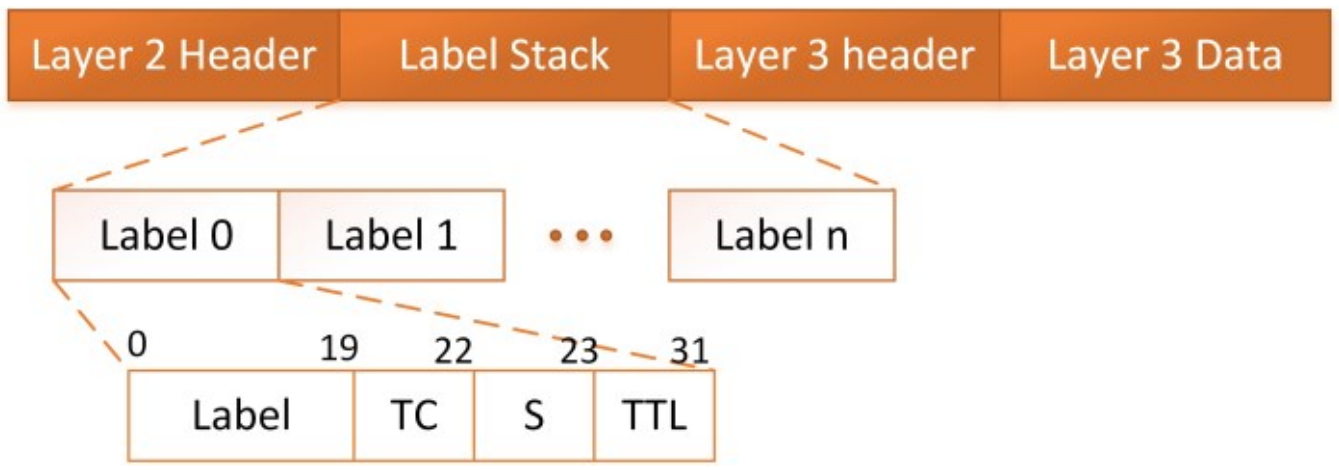

Fig.4. MPLS Label stack and Label format.

- Segment routing over MPLS data-plane: Segment routing can be directly applied to the MPLS data-plane and this does not require any change in the MPLS forwarding plane, as labels are now used as segment identifiers while the segment header is instantiated as the MPLS label stack [17]. There is no requirement for any modification in the signaling protocol to advertise the labels. Globally unique identifiers are assigned from a segment routing global block (SRGB) [18] and local identifiers are assigned as a combination of SRGB and a constant prefix. A SRGB is a range of label values reserved in label distribution database of a segment routing enabled router for global segment identifiers. With MPLS, the *continue* instruction of the SR technique now corresponds to *swap* operation in MPLS. The MPLS label structure and MPLS label stack is shown in Fig. 4, which can be used for SR without any major modification.

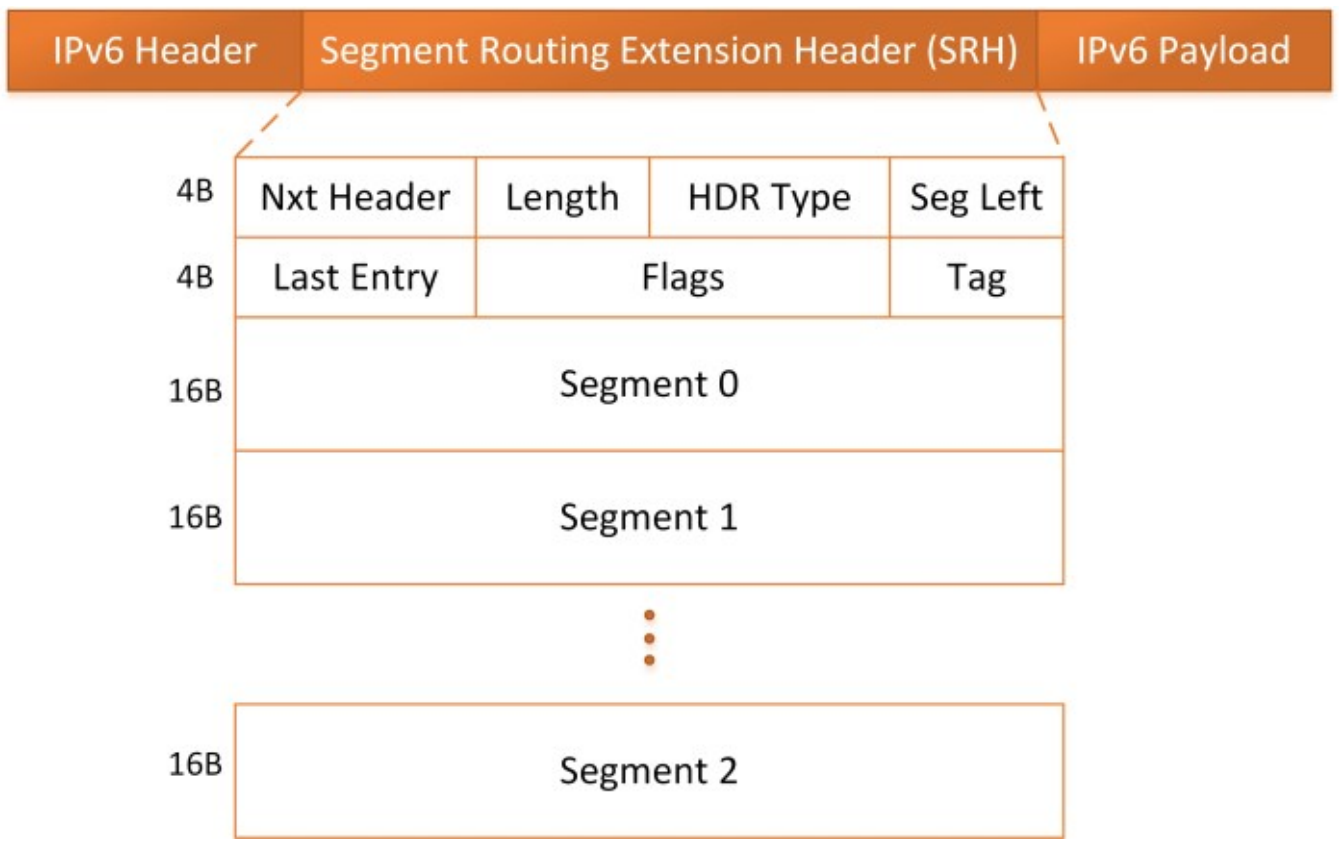

Fig. 5. Segment routing header extension for IPv6.

- Segment routing for IPv6 data-plane: In the absence of a MPLS data-plane, segment routing in the network is proposed by using the IPv6 data-plane known as SRv6 [19], [20]. SRv6 defines the SID as a 128-bit IPv6 address, which makes the scheme simpler from the signaling perspective. By making the segment identifier as the IP address itself, the scheme actually becomes simpler to implement than the vanilla MPLS case. A node need not advertise any new identifier other than an IPv6 address. A new routing header extension is defined in IPv6 to include the segment list in the header now known as the segment routing header (SRH) [21]. SRH consists of a segment list, a pointer that points to the active segment (i.e. the segment to act upon), and other optional fields. The SRH header format is as shown in Fig. 5.
For the forwarding of a packet an active segment identifier (IPv6 address) is placed in the IP DA field of the IPv6 header. All intermediate routers forward the packet based on the IPv6 destination address (DA) without inspecting the SRH header. Therefore, this process also provides interoperability between an SR-capable and non-SR-capable node. Once the packet reaches its destination specified by the IPv6 DA, the SRH header is inspected for the next segment. In case, this was the last segment, the SR header is removed and packet is forwarded in its native format. Otherwise, the *segment left* field is decremented and the DA field of the IPv6 header is replaced by the next SID in the list.
- Customized data-plane: A customized data-plane can be created by defining a new protocol such as the one shown in [22], [23]. This new protocol has to then define the SR header format and its encapsulation/decapsulation in the packet. The new protocol also needs to define the segment identifiers and their encoding scheme in a packet. For example, segment identifiers in [22], [23] are generated by creating auxiliary graphs of nodes for a router. Each node in the auxiliary graph has up to three edges, one incoming and two outgoing. Based on the outgoing edge whether it is towards the left or right of the incoming edge a 0 or a 1 is assigned. The number of bits in an identifier is equal to number of auxiliary nodes traversed by a packet. This way an identifier is generated for each of the interface of the router. This generated identifier is similar to the Adj-SID of the originally proposed SR architecture [19]. Usually customized data-planes use a centralized controller for the path computation and generation of segment list, which is further used for the configuration of an individual node.

The comparison of different data-planes for SR is presented in Table I.

### *B. Segment routing control-plane*

The control-plane of the SR framework defines: a) the distribution of the labels to SR enabled nodes; and b) the selection of segment-routed paths. We now discuss these two aspects of the SR control plane in detail.

#### *1) SR Label distribution*

For proper working of segment routing, the exchange of SID information among the nodes and evaluation of shortest path between the nodes in the network is important. The SR-control plane allows the exchange of this information using link state IGP protocols such as IS-IS and OSPF, and their extensions [24], [25], [27] [28]. Extensions to link-state IGP protocols are defined to allow a node to store the SIDs in their database. We discuss different protocols and their extensions to support segment routing in appendix.

#### *2) SR path computation*

The SR control plane instructs the node to select an SR-path for a packet/service. One or more of the following methods are used for path selection:

*a) Distributed Constrained SPF based selection*

In this method, a node selects a path based on some policy constraint i.e. link utilization, bandwidth, latency etc. After selecting the path, a list of nodes and adjacencies are computed for the path encoding. Different path encoding strategies and algorithms are discussed in the next section.

*b) SDN controller-based selection*

In this method, a centralized SDN controller is used to compute a path and encode the computed path into a list of segments based on different available algorithms. This scheme provides flexibility in computation of the traffic-engineered path because of the availability of network-wide link utilization statistics.

*c) Statically defined by an operator*

In this method, static paths are defined by the operator. This method is useful for testing and troubleshooting purposes.

## III. SR LABEL ENCODING SCHEMES

SR does not require any signaling protocol for label stacking. However, this may introduce scalability and packet overhead issues due to addition of segment routing header in a packet. Segment routing can be provisioned by adding a segment for each node along a path in a segment list, but forwarding of a packet to its destination could be done with a segment list of smaller size. For example, in Fig. 6 a packet from node-*A* to node-*G* needs to follow the path *A-B-E-G*. A segment corresponding to each node of the path could be added in the segment list for a packet to be forwarded. This results in a four-segment path. Since the shortest path from node-*B* to node-*G* passes through node-*E*, therefore the segment representing node-*E* in the segment list can be removed. As any packet going from node-*B* to node-*G* over a shortest path must go through node-*E*. As a result, we now have only three segments in the final list as opposed to four segments we had previously. Therefore, specific algorithms are required to calculate the optimum label for a given path. This section discusses different label encoding algorithms used in the SR.

Segment routing based labeling schemes can be of two types: 1) Strict-routed; and 2) Loosely-routed. A SR-path is said to be strict routed, if a segment representing each of the node/link of the path is present in the SR header. On the contrary, a path is said to be loosely routed if the SR header contains a number of segments that are fewer than the nodes/links over the path. Strict encoding of an SR-path generates list of segments of maximum size to encode the path in the packet. This type of encoding is essential for testing as well as for operation and management tasks. However, this scheme cannot be used to forward the data traffic in a service provider network. Using strict encoding may not be feasible due to the violation of the maximum segment depth constraint (Maximum permissible segments in a packet). In this section, we discuss different label encoding algorithms proposed to effectively utilize the segment routing scheme to minimize the list of segments and also to optimize the selected paths.

In [29], two algorithms SR-D (Direct) and SR-R (Reverse) are presented. SR-D traverses along the path from the source to the destination node while SR-R traverses from the destination to the source node. A centralized SDN controller is considered to be associated with a segment-routed network. Whenever a new flow has to be established, the SDN controller is requested to establish the path for a traffic flow. Both algorithms work on this precomputed path and use an iterative method to navigate the target along the computed path to generate a segment list.

Both algorithms start with the initial two nodes of the path, subsequently nodes along the path are added iteratively. If the path is a unique shortest path, then the algorithm adds the next node in its targeted path and takes the previous node SID as the segment, otherwise it breaks the loop at the node. If the loop has been broken before reaching the last node of the path, then the algorithm builds a new target path starting from the node where the last path was broken. For each break in the path a segment is added in the list. Once a full path is navigated a list of segments is generated which can be encoded in the packet to reach the destination node.

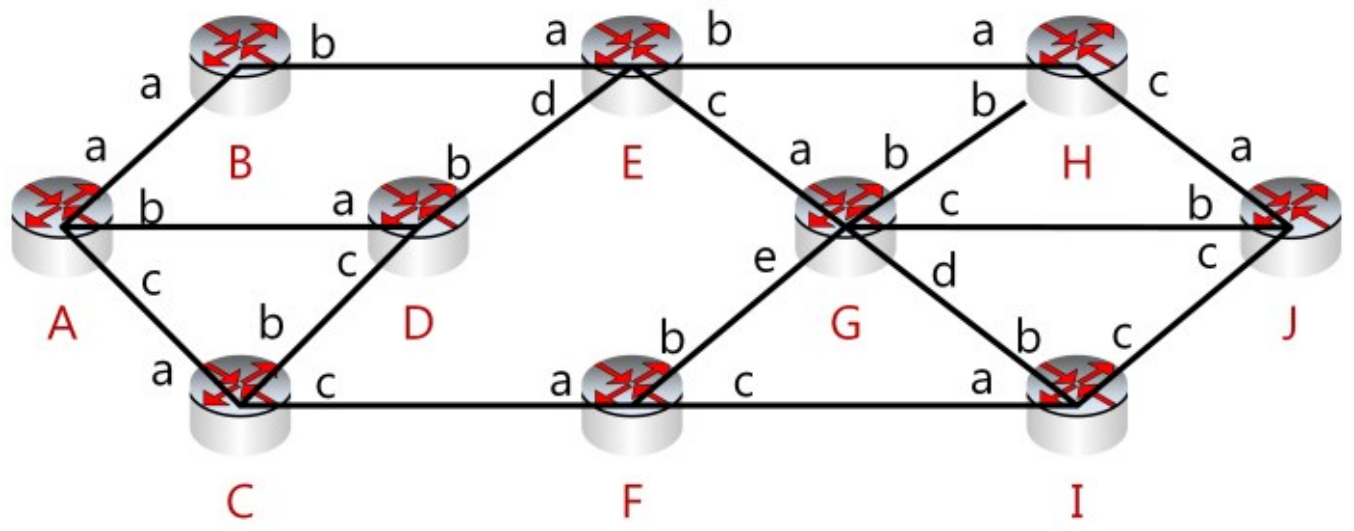

Fig. 6. Segment routing illustration.

For example, consider the topology in Fig. 6 which has three shortest paths from node-*A* to node-*G*. Let us assume that the controller computes the path *A-D-E-G* for the segment list generation. The algorithm then starts with the initial two nodes of the path, i.e. node-*A* and node-*D*. Since the path between node-*A* and node-*G* is not a unique shortest path, therefore when we iterate over the path, then the iteration breaks at node-*D*. Subsequently node-*D* is added in the segment list. A new iteration starts for the path between node-*D* and node-*G*. Since we are again considering the initial two nodes, this new iteration now starts with node-*D* and node-*E*. Since, node-*E* is on the shortest path to the destination node-*G,* therefore, the next node, i.e. node-*G* is added iteratively for segment computation. Node-*G* is in this case the destination for this path and therefore a second segment *G* is added in the segment list and the loop breaks. This generated segment list *{D, G}* is used for segment routing along the path *A-D-E-G*.

SR-R works similar to SR-D except that SR-R adds the start-node in segment list instead of the last-node and once a segment list is generated, the list is reversed at the final step. For example, let us again consider the path *(A-D-E-G)* from the above illustration. SR-R starts its iteration from node-*G* and node-*E* and generates the segment list as *{G, D}*. This generated list is then reversed in the final step and provides a list *{D, G}* to be used at node-*A* for forwarding. Both algorithms compute the segments effectively and generate the segment lists with minimum depth.

In [30] two algorithms (SR-LEA – Segment Routing Label Encoding Algorithm and SR-LEA-A – Segment Routing Label Encoding Algorithm-Adjacency SIDs) for encoding the segment routed path are presented. The performance of the algorithms is benchmarked over several real-world network topologies such as GEANT, Germany50 etc. Results show that proposed algorithms efficiently reduce the size of segment list.

The SR-LEA algorithm takes the precomputed path as an input to generate a list of segments to be encoded in the packet. The precomputed path can be calculated manually or by a centralized SDN controller or by a PCEP element. The pre-computed path is then broken into sub-paths, with each sub-path comprising of two or three nodes. For a three node sub-path, the node-SID of the last node is added as a segment in the list, which is then used by a packet to traverse the sub-path. For two node sub-paths, the Adj-SID of the interface connecting the two nodes is added to the segment list for a packet to traverse from one sub-path to another. For example, in Fig. 6, to traverse from node-*A* to node-*J,* the controller computes a path *A-C-F-I-J.* The algorithm takes this computed path *(A-C-F-I-J)* and breaks it into two sub-paths *A-C-F* and *I-J.* After partitioning, the segment is computed for the individual sub-path. Node-*A* has a unique shortest path to node-*F* through node-*C,* which implies that a segment *F* is thus generated for the first sub-path. The second sub-path has only two nodes, therefore segment *J* is generated for this second sub-path. A third segment using Adj-SID is also added for joining two sub-paths, i.e., *c*. The final segment list generated by the algorithm is then denoted by *{F, c, J}*. All generated SIDs are sequentially added in the label. This algorithm reduces the size of segment list by 52-67% compared to the strict routing case.

The SR-LEA-A algorithm is similar to the SR-LEA except that it utilizes Adj-SIDs, which are globally unique. By utilizing the global Adj-SID, the size of segment list can be further reduced. To traverse an interface represented by a local Adj-SID, a packet needs to have a Node-SID as well as a local Adj-SID. The Node-SID enables the packet to reach the specified node and then the Adj-SID facilitates to identify the particular interface. However, in case of global Adj-SID, the requirement of Node-SID can be eliminated. As a result, the depth of the segment list can be further reduced. For instance the segment list generated by the SR-LEA in the above example is *{F,c,J}* and this can be reduced to *{c, J}* by SR-LEA-A algorithm. This algorithm reduces the size of segments list by 57-67% compared to the strict routing case.

TABLE II
ILLUSTRATION OF TRAFFIC SPLIT FRACTIONS

| $(i,j)\backslash k$ | $B$ | $D$ | $E$ | $F$ | $G$ | $H$ | $I$ |
|---|---|---|---|---|---|---|---|
| $(A,J)$ | 0 | 0 | 0.3 | 0.7 | 0 | 0 | 0 |
| $(A,H)$ | 0.5 | 0.3 | 0 | 0.2 | 0 | 0 | 0 |
| $(D,J)$ | 0 | 0 | 0 | 0 | 0.4 | 0.2 | 0.4 |
| $(C,E)$ | 0.5 | 0.4 | 0 | 0 | 0.1 | 0 | 0 |

*Traffic Engineering and Segment Routing:* Algorithms are presented in [31] to optimize the segment routed paths. An algorithm that is oblivious to the traffic matrix is presented for the offline segment computation and another competitive algorithm is presented for online segment routing. Both algorithms focus on the concept of *2-segment routing*, where traffic between the source and destination nodes pass through exactly one intermediate node. In 2-segment routing, the key decision is to pick appropriate segments for each flow in order to minimize the overall network congestion. The main difference between the offline and online algorithms is that for the offline case, traffic split parameters are used. The split parameters are defined as the fraction of traffic passing through a node for a source-destination pair. The split parameters are computed ahead in time and are setup once. Whereas for the online case, segments are chosen based on the current link utilization, which requires continuous monitoring of the links. We now describe both of the algorithms.

*Offline algorithm*: A rigorous approach based on a game theoretic technique is used in this algorithm. A fraction of traffic $a_{ij}^k$ that passes through an intermediate node-*k* for a source-destination pair $(i,j)$ is set and which is independent of the traffic matrix. This fraction of traffic $a_{ij}^k$ is called the traffic split variable. This traffic split variable is calculated using a linear programing formulation. The objective of the linear programing model is to minimize the maximum link utilization by choosing suitable traffic split variables such that the traffic is distributed over different paths. The traffic split variables are constrained such that $\Sigma_k\, a_{ij}^k = 1$ . This ensures that the sum of split variables for a pair $(i,j)$ over all the intermediate nodes should be 1. Once the traffic split variables are selected, a hash-table is instantiated at each ingress node for a particular path. This hash function takes the flow- ID as input and returns a *k* value, this *k* indicates the intermediate node for the flow to pass through. A two-segment list is created by using the node-SID of the node-*k* (an intermediate node) and node-*j* (destination node).

Table II shows an example of traffic split values for different source-destination pairs $(i,j)$ over the network topology as shown in Fig. 7. For a source-destination pair $(A,J)$, if we choose node *E* as the intermediate node *k*, then this node shall carry 30% of the traffic. Similarly, for a source-destination pair of $(C,E)$, if node-*B* is chosen as an intermediate node, then it will carry 50% of the traffic.

*Online algorithm*: This algorithm is beneficial when the network has a central controller like PCE or SDN that processes all the flow requests and computes a segment-routed path. The objective of the online approach is to accommodate as many flows as possible without rejecting requests. In this approach, a central controller determines the intermediate node-*k* to route the request. It is assumed that the controller knows the status of all the links and the nodes in the network. Upon arrival of a traffic request, the controller uses state information to compute a path and determines an intermediate node. Performance evaluation shows that the online algorithm works better than the shortest path segment-routing approach with the rejection rate being low due to congestion avoidance.

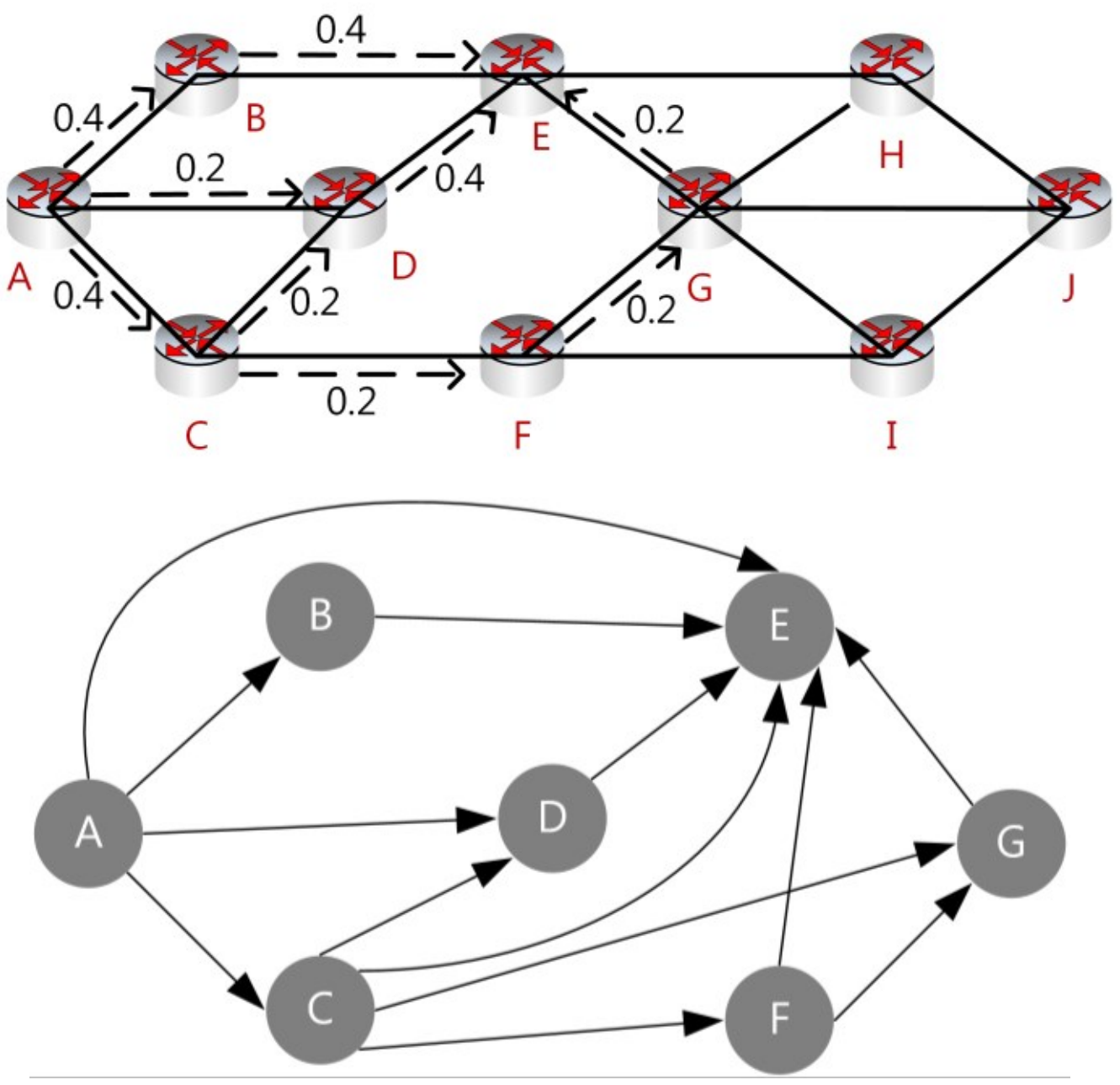

Fig. 7. An example of optimized path between source node-A and destination node-E (Top). Auxiliary graph based on the optimized path (Bottom).

*Segment Routing for Traffic Engineered Paths:* A traffic engineering-based segment encoding scheme is presented in [32]. It is considered that a set of forwarding paths *P*, are configured in the network by a network operator based on a set of given policies. This work also considers multipaths for forwarding. A forward splitting vector is also considered that indicates the fraction of traffic between a source and destination flowing on a particular link for a forwarding path. This vector is generally zero/one for single path forwarding. Based on the available forwarding paths, the network operator computes the optimized path using traffic engineering (TE) algorithms for every source-destination pair. The segment encoding problem is composed as a two-step process: 1) the creation of an auxiliary graph and; 2) the solution of a Multi-Commodity Flow (MCF) over the auxiliary graph. These steps are described in following sub-section.

*1) Building the auxiliary graph*: In this step, an auxiliary graph is created based on the forwarding paths and the optimized paths. The algorithm takes the network graph, forwarding path and the optimized path as inputs, and produces an auxiliary graph as the output. An example of an optimized path between a source-destination pair *A-E* is as shown in Fig. 7. The auxiliary graph is created for a source-destination pair based on a traffic request. This graph contains only those nodes which are on the optimized path. There are a few additional links added between the nodes in an auxiliary graph by the algorithm. A direct link between the nodes are added in case all the forwarding paths between two nodes are also part of the optimized path. An example of the auxiliary graph creation for a source-destination pair *A-E* is as shown in Fig. 7. In the auxiliary graph there is a direct edge between *A-E* and *C-E*. This implies that all optimized paths from *A* to *E* are present in the graph. Similarly, for edge *C-E* all the optimized paths between node-*C* and node-*E* are also present.

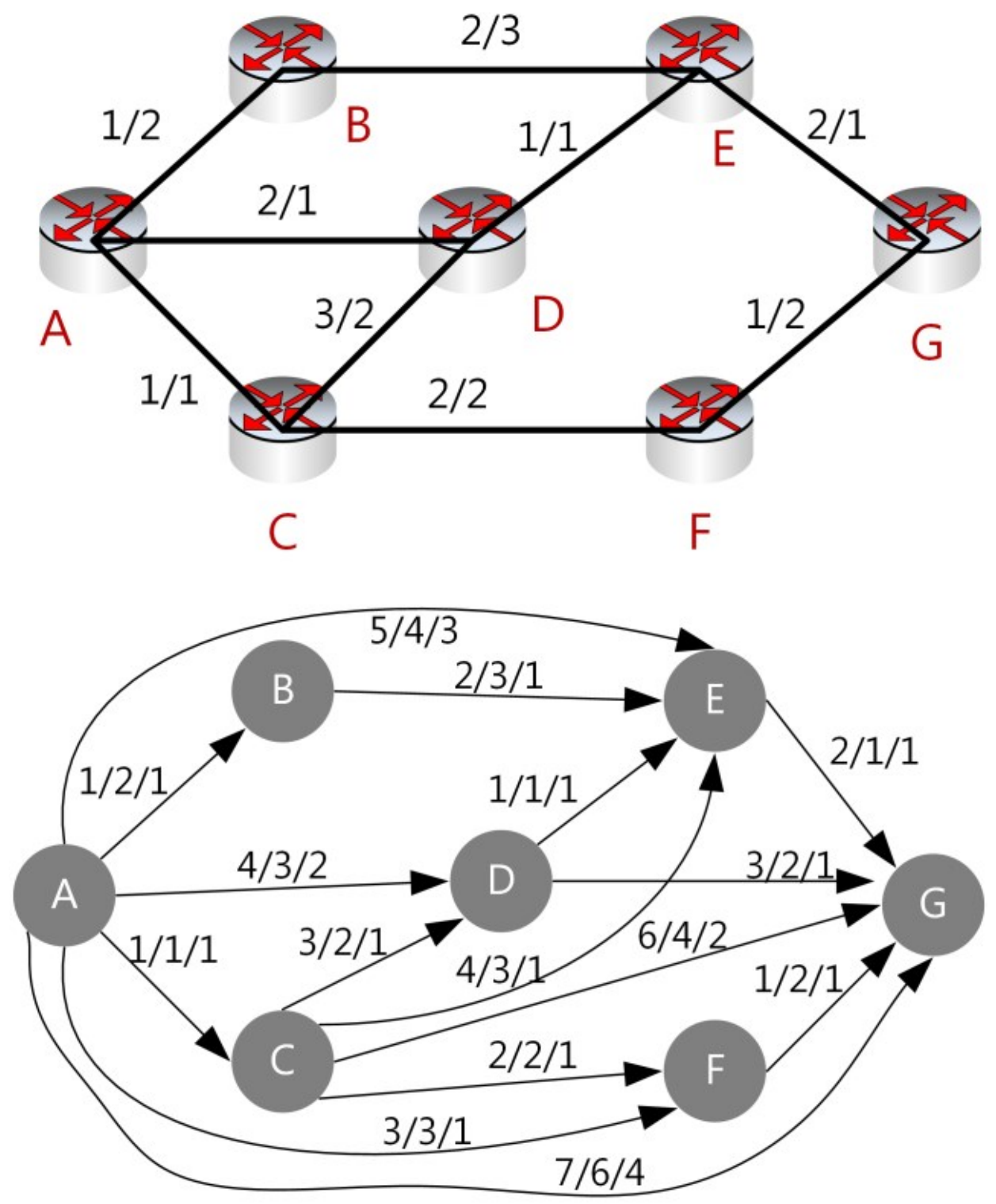

Fig. 1. Network graph G (Top), each link is associated with IGP cost/Latency. Auxiliary graph (Bottom), each link is marked by (worst case IGP cost)/(worst case latency)/(ECMP path)

*2) Multi-commodity flow problem*: This step of label encoding is presented as a MILP formulation, where the objective is to minimize the total length of the segment list to be used. The MILP uses an auxiliary graph for computation of the segment list. Each path in the auxiliary graph from the source to the destination represents a SR path. There are multiple options to navigate from the source to the destination in an auxiliary graph. The proposed method helps in minimizing the number of SIDs composed in a segment list required for forwarding and also leads to reduction of the overhead.

The authors in [32] evaluate the performance of the algorithm over different topologies using LSP tunnels, which are defined explicitly similar to strict routing. Using explicit encoding, it was observed that 50% of the paths had segment size greater than 7 segments. In contrary, 90% of the paths provided by the algorithm described above had segment size less than 4 segments. The maximum size of the segments when explicit encoding was deployed was 13, whereas with the proposed algorithm this size was reduced to 7 segments. As a result, there was a significant reduction in the packet overhead.

An algorithm in [33] was proposed for the computation of the ECMP-aware SR paths satisfying the constraints such as inclusion/exclusion of specific node(s)/link(s), latency, IGP-metric etc. The algorithm considers the available shortest path between any node-pair. The shortest available path can be provided by a centralized $E$ and $V$ are edges and vertices in Network graph. $E'$ and $V'$are edges and vertices in auxiliary graph. $d$—average path length.

TABLE III
COMPARISON OF THE DIFFERENT LABEL ENCODING SCHEMES

| PARAMETERS | SR-D/R [29] | SR-LEA/-A [30] | TRAFFIC OBLIVIOUS/ONLINE [31] | MCF BASED [32] | ECMP AWARE [33] | OE [22] |
|---|---|---|---|---|---|---|
| Data-plane Support | All data-planes | Proposed for MPLS data-plane but can be extended to any data-plane. | All data-planes | Good for MPLS but can be used with others | All data-planes. | Customized data-plane |
| Header Overhead | Low | High | Low | Moderate | Moderate | Moderate—vary with the path length |
| Path selection scheme | Shortest path | Traffic engineered path | Traffic engineered path | Traffic engineered path | Traffic engineered path | Traffic engineered path |
| Traffic engineering capability | Low—due to shortest path selection. | High—due to path partitioning. | Low—only two segments are used for routing. | High—due to multi-commodity flow solution. | Moderate—path length greater than a threshold are rejected. | High—select the best available path based on the rate limiting and other policies. |
| Path computation element | Central controller | Manual/Centralized controller/ PCEP | Manual/Centralized controller/ PCEP | Network operator | Centralized controller/ Network operator | Centralized controller |
| Congestion over links | High | Low | Moderate | Low | Moderate | Low |
| SID used | Node SID | Node SID and Adj-SID | Node SID | Node SID | Adj-SID and Node SID | Similar to Adj-SID |
| Complexity | $O(E+VlogV)+O(d)$ | $O(E+VlogV)+d/3.O(d/3)$ | $O(V^3)+O(V)$ | $O(V^3)+O(E'+V'logV')$ | $O(V^3)$ | $O(E+VlogV)$ |
| Approach | Iterative approach—Path computation is done by the controller. Iteration over the computed path for the segment list generation starts from source/destination node. | Path Partitioning-Path is computed by the PCE. Computed path is partitioned in sub-paths of length two or three nodes. | Two segment routing—Path is computed using the path computation element. Path is broken into two sub-paths. One segment is generated for each sub-path. | Constraint optimization model based on multi-commodity flow—traffic is splitted into multiple flows and different paths are assigned to different flows. Objective of the problem is to minimize the length of segment list and accommodating the maximum flows. | Policy based—paths are computed based on the set of policies/rules and exclusion/inclusion of links. ECMP aware paths are considered. | Binary routing—creates the auxiliary graph of the nodes and follow the binary route. This binary route is the segment list. |

SDN controller or manually configured by a network operator. The proposed algorithm starts with the computation of the path satisfying the specific set of policies, rules or some other pre-decided metric, such as latency and inclusion/exclusion of a specific node etc. Once the path computation is complete, a network graph *G* is constructed such that each link of the graph is associated with the two parameters: IGP path cost and latency.

An example of a network graph is shown in Fig. 8 (Top). For each pair of nodes in *G*, new virtual links are added, thus representing the set of ECMPs between two nodes. The virtual links have three parameters associated with each link *x/y/z*, where $x$ is the worst case IGP path cost, $y$ is the worst-case delay between nodes, and $z$ is the number of ECMP paths between the nodes. An example of an auxiliary graph is shown in Fig. 8 (Bottom). The paths between a source-destination pair are again computed over the auxiliary graph, where the paths having a number of hops greater than some preset threshold are rejected. The remaining paths are sorted based on path-length and associated parameters *x/y/z*. Resulting paths are the combination of physical and virtual links. For the generation of the segment list, physical links are mapped to the adjacency SIDs and virtual links are mapped to the node SID of the last node along the path.

In [34], a TE/SR based heuristic is presented for flow allocation. The heuristic is implemented in an SDN controller. This controller uses the traffic flow-rate and link bandwidth information to compute a path for the traffic flow. The computed path is then used for the generation of a segment list for an SR path. The proposed heuristic is divided into two steps: 1) a constraint shortest path first (CSPF) phase and 2) a re-assignment phase. In the first phase, allocations of the flows are realized. In case a flow cannot be realized in the network across any path, then that flow is rejected. In the second phase, all the realized flows are re-assigned one-by-one so as to minimize the average latency. The second phase is executed multiple times until a segment list of the smallest depth is generated. The second phase iteration terminates when there is no improvement in the generated segment list size. The heuristic is implemented and experimented on a small testbed network. A comparison is performed for the distribution of path-length between a traffic-engineered path and the shortest-path. Results show that the path-length calculated by the heuristic is comparable to the shortest-path length.

In [22], traffic engineered paths are calculated between a source-destination pair based on link-utilization and bandwidth constraints. The work in [22] uses a strict routing scheme, where the list of segments in a packet has a SID for each link to be traversed by the packet. Due to the use of strict routing, this scheme provides traffic-engineering capability at a larger scale but with a trade-off – that of incurring a large packet overhead. To overcome the large overhead, a different scheme is proposed in [23], where the authors fixed the maximum length of the header that can be inserted in a packet. In case the provisioned path leads to a situation of the header length exceeding the maximum length, the path is then broken into sub-paths and intermediate routers are selected based on the load. These intermediate routers are termed as the *multi-segment supporting nodes*. The process of breaking a path into sub-paths helps in reducing the packet overhead. The ingress node inserts the list of segments to forward a packet to a multi-segment supporting node. Once a packet reaches a multi-segment supporting node, the SR header in the packet is swapped by a new SR header. This new SR header contains a list of segments that leads the packet towards the next multi-segment/destination node. This scheme helps in reducing the overhead significantly, but this also requires a table lookup at multi-segment nodes.

In Table III, we compared various encoding schemes based on different parameters. All the encoding schemes are carried out for a particular data-plane implementation but can be extended to other data-planes as well, with minor modification. We compared the header overhead with strict SR, where a SID representing each of the nodes on a path is present in the segment list. Traffic engineering capability is analyzed based on the scheme's capability to utilize the different paths in case of congestion. We also compared the complexity of the algorithm proposed for the encoding scheme.

## IV. Segment Routing Use-Cases

Segment routing has strong potential to be utilized for variety of applications such as restoration, load-balancing, monitoring, as an SDN data-plane, network function virtualization and others. We cover in this survey paper some of these use cases and note that with the growth of segment routing newer use cases shall appear on the horizon. In this section, we discuss some of the important use-cases, where the application of segment routing has been successfully demonstrated.

### A. *Restoration*

An important application of segment routing is the automatic re-routing of connections post a failure. Re-routing can be done with available IGP based mechanisms or by the other alternate mechanisms such as loop-free fast re-route. The main challenge in segment routing-based restoration is the identification of the primary path such that the restoration bandwidth of the protection path is shared by different flows that potentially fail at different time instances. A solution to the problem is presented as a linear programming formulation in [35] and an efficient primal-dual algorithm is developed.

In the presented solution [35], it is assumed that the network is controlled by a centralized SDN controller, which has full knowledge of the network topology and traffic requests. The controller uses this information to compute the SR path for each source-destination pair and updates the path information in the form of segment lists at an ingress node. Whenever, there is a failure in the network, IGP recomputes the shortest path without the intervention of the controller and new routes are restored once IGP converges.

A mathematical model is developed to ensure the spread of traffic between different paths. The network is represented as a network graph and each edge of the graph is associated with an IGP link weight and corresponding capacity. It is assumed that traffic between a source-destination pair can split over multiple paths. The splitting of traffic is done at the flow level to ensure that all the packets of a particular traffic request follow the same path. The following scenarios are considered.

#### 1) *Restoration planning for a single link failure*

It is assumed that IGP weights of the links are given for a network. This implies that the shortest path between a source and destination is fix. A 2-segment routing scheme, as described in section III, is used to reduce the computational complexity. In the two-segment routing scheme, an intermediate node is picked for every traffic request to pass-through for a source-destination pair. In case an intermediate node is already on the shortest path then the destination node SID is used as the segment, otherwise the node SID of the intermediate node followed by the SID of the destination node are used to generate segment list. The objective of the controller is to pick the segments in a manner such that sufficient bandwidth is available to handle the traffic. An optimization problem is formulated with the objective to minimize the maximum link utilization. Link utilization includes the primary traffic as well as the restored traffic that increased due to the link failure. A restoration bandwidth equal to the maximum traffic required to be restored is reserved on the link for a link failure scenario. Since, link failures are assumed to be independent events, therefore enough bandwidth to carry the traffic in the worst-case should be reserved. Instead of solving the problem of minimizing the maximum link utilization, we maximize the inverse of maximum link utilization. This results in an objective of maximizing the throughput. A primal-dual algorithm is developed to solve the problem in polynomial time.

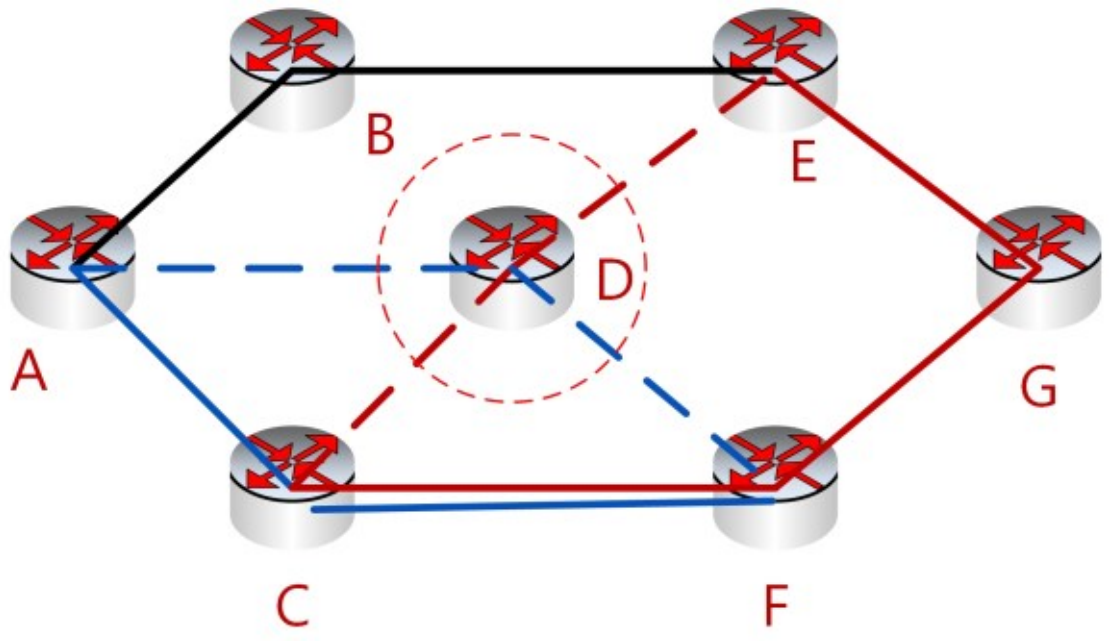

Fig. 9. Illustrating SRLG and impact of resource failure on the routing. Links AD, CD, FD and ED forms a SRLG. Dashed line represents the flow before the failure and solid lines (red and blue) shows the flow after failure.

### 2) *Shared risk link group failure*

A SRLG is a set of links sharing a common resource such that failure of the resource leads to failure of all the links sharing the resource [36]. Unlike a single link failure, SRLG may require a change in its segment header as a new segment list is to be created due to the association of failed resource in the previous segment list. This may involve changing the list of segments at the ingress node. Therefore, a new path is computed after removing the failed resource and associated node(s)/link(s) from the network. For example, in Fig. 9, node-*D* is traversed by two flows: one from node-*A* to node-*F* (blue dashed line) having segment list *{D,F}* and; another from node-*C* to node-*E* (red dashed line) having segment list *{D,E}*. The set of links *{AD, CD, FD, ED}* form a SRLG, where a failure of node-*D* results in failure of all the links mentioned above. Upon failure of node-*D*, the two paths going through node-*D* also fail. This implies loss of the segment list. Hence, in such a scenario, it is required to change the segment header. Failed links and nodes are removed from the network and a new segment list *{C,F}* and *{G,E}* indicating the new paths are computed for both the flows.

### 3) *Equal Cost Multiple Paths (ECMP)*

In the case of provisioning ECMP, traffic is split between multiple equal cost paths. The splitting is usually done based on a hash on the flow-ID and all the packets belonging to same flow thereafter follow the same path. For ECMP, the fraction of traffic between any source-destination pair flowing through a link is computed. This (fraction) information is used to formulate the traffic restoration problem.

A comparison between the shortest path routing and two segment routing is evaluated without restoration in [35]. The authors in [35] conclude that two-segment routing performs similar to shortest-path routing. With restoration, segment routing shows enormous bandwidth efficiency in comparison to shortest-path routing. The authors also demonstrated that segment routing resulted in better throughput compared to shared local restoration.

## *B. Traffic Engineering for Energy-Efficient Backbone Networks using SDN*

In [37], an SDN based approach is presented to improve energy-efficiency in a backbone core network. The segment routing strategy is used for adapting dynamic traffic by selectively turning ON/OFF a subset of links (based on link utilization) in the network. Links cannot be simply turned off without losing data. The work in [38] has shown that even micro-second sleep times are difficult to achieve due to small packet arrival gap. Earlier works have shown to reduce link utilization by changing the path of a flow and making the link available to incorporate sleep time [39]. A centralized approach is presented in [40], where MPLS+RSVP-TE is used to estimate traffic matrices. In this work, segment routing is used, which allows similar benefits compared to the work presented in [41]. This work also leads to a reduction in protocol complexity. The key assumption in this work is that variations in the traffic pattern in the backbone network is slow. The solution presented in [37] is achieved in three steps:

- Selecting links to switch ON/OFF: In this step, suitable links that can be turned ON or OFF are identified. For the identification of such a link, we use the “least congested link” technique. For this purpose, all the links are sorted based on link-utilizations. Each link is examined for a connectivity constraint, in order to ensure that all the nodes are connected, even after the identified links are turned OFF. This step also identifies which link to turn ON as the traffic demand grows between a source-destination node-pair and the link begins to get congested.
- Computing new routes: Once a link selected to turn OFF/ON is identified, new routes are calculated for the traffic by excluding/including the selected link. The SDN controller exploits its network-wide knowledge as well as the link utilization availability to compute new routes for the flows.
- Rerouting and switching links OFF/ON: Once the new routes are computed, the controller writes the new configurations for the affected traffic flows at the ingress node. The controller also sends messages to nodes that are connected by the link selected for turning OFF/ON.

The SDN controller continuously performs the above steps in a loop fashion to compute which links to turn ON/OFF. The presented use-case was implemented in the OMNET++ simulator and shown in [37]. This solution was tested over two network topologies with real traffic matrices from SDNLib for a period of a day. Results shows that on an average ~6KWh of energy consumption was reduced. Due to rerouting and use of the traffic engineered path the authors report that there was an increased overhead of ~18%. Also, a 20% average increase in delay was shown due to variation from the shortest path.

## *C. Network Function Chaining*

Network Function Virtualization (NFV) [42], is an emerging paradigm in which network functions are moved from traditional hardware to software that runs on virtual machines in commodity servers. NFV facilitates creation of service function chains (SFC) [43] in the network that are an ordered set of virtual network functions (VNFs) that a packet must traverse. The VNFs can be at various layers of the protocol stack such as firewalls, load balancers, TCP optimizers, deep packet inspection, etc. The SFC architecture is defined in [44]. A new packet header to support the SFC of VNFs is proposed in [45]. This packet header contains the list of VNFs to follow. In [46], a segment routing approach is used to create VNF chaining where the list of segments in the header decides the order of VNFs to follow. VNFs process data by removing the SR header and the same/modified header is inserted again in the packet once the processing is complete to forward the packet to another SR aware VNF.

An architecture is proposed based on IPv6 SR where the nodes in the network are of three types: 1) traditional IPv6 routers, 2) IPv6-based SR enabled routers and 3) NFV nodes. NFV nodes are capable of hosting VNFs. Such types of nodes can be IPv6-SR nodes with additional functionality to host the VNFs.

In [46], SRv6 is exploited and the IPv6 address of the node is considered as the SID. Further, the list of segments in the header is considered as the VNF chain. At the ingress node, the original IPv6 packet is encapsulated with the SR header as the payload in the IPv6 packet. The new IPv6 packet has the ingress node IP as the source IP address and the next VNF address as the destination address. As this encapsulated packet arrives at an NFV node with the destination address same as that of the node, the packet is then processed and forwarded to the designated VNF. Once the packet is processed by the VNF, the packet is then passed to the NFV node for further forwarding.

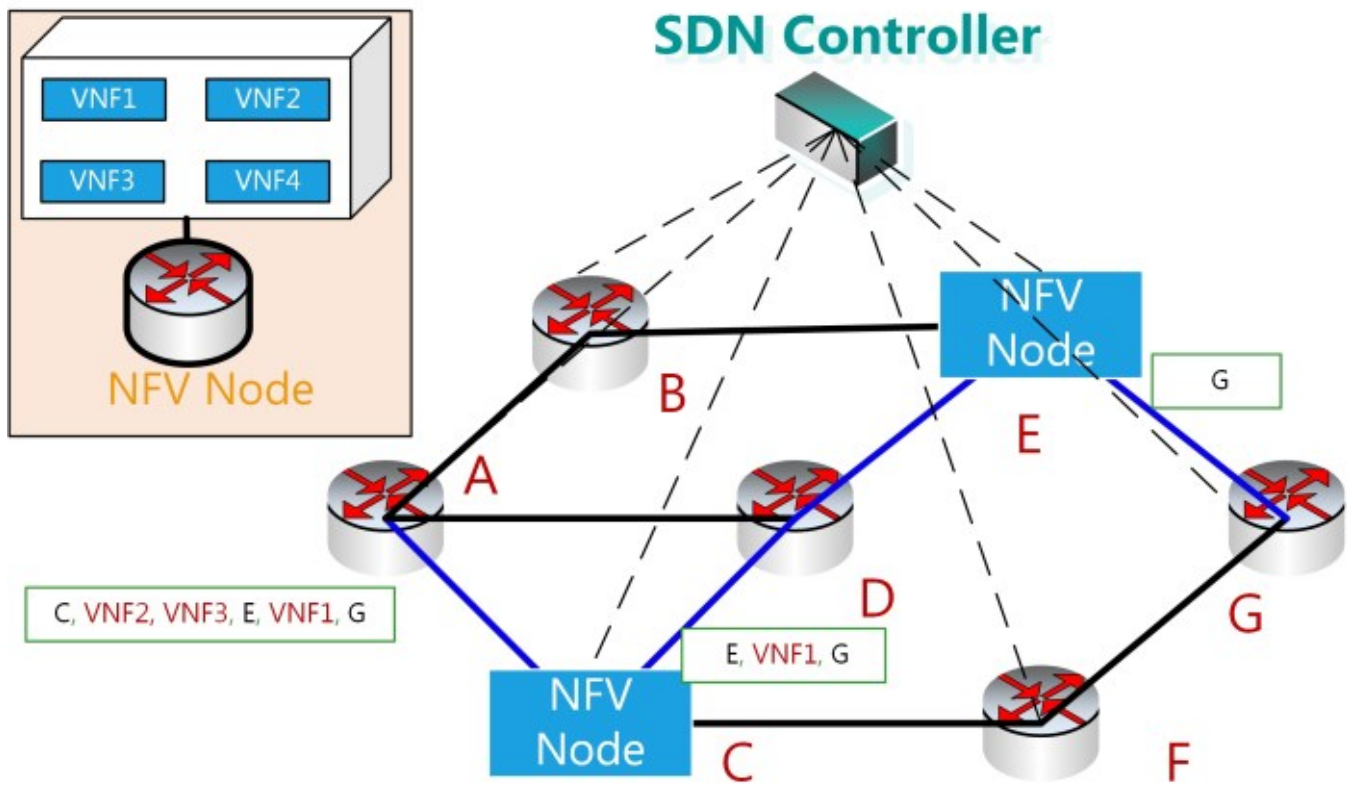

Fig. 10. Service function chaining using segment routing. NFV node is shown in inset. A SR path is shown by the blue line. The list of segments to be inserted at node-A is shown in green rectangle.

An illustration of service function chaining using segment routing is shown in Fig. 10. The figure also shows the architecture of an NFV node, whereby the NFV node consists of a SRv6 [19, 20] node and server(s) to host the VNFs. NFV node-C hosts two VNFs: *VNF2* and *VNF3*, whereas NFV node-*E* host only *VNF1*. For a traffic request from node-*A* to node-*G*, this request requires to be processed by the VNFs in the sequence *VNF2, VNF3, VNF1*, thereby creating a service chain. A SR path is set from node-*A* to node-*G* through NFV nodes such that all the VNFs part of the service chain are present along the path. The sequence of VNFs to be processed is added in the segment list as a service SID (green box in node-*A*), which are locally significant to a node. Once a SR encoded packet reaches the NFV capable node-*C,* the packet is then passed to the server hosting *VNF2* by the SRv6 part of the NFV node. Thereafter, the packet is processed by *VNF2* which forwards the packet to the SRv6 part of the node by making the next segment (*VNF3*) active in the header. Subsequently, the SRv6 part of the NFV node forwards the packet to the server hosting *VNF3*. Once all the VNFs to be processed by the NFV node are traversed, then the packet is forwarded to node-*E* by SRv6 part of the NFV node.

The control plane of the architecture includes an orchestrator that interacts with the NFV managers for the purpose of administration and configuration. The control plane also has an SDN controller for configuring the nodes in the network. Each NFV node is associated with a SR/VNF connector module for forwarding the incoming SR encoded packets to the respective VNFs. This operation is performed in three stages: i) identification of the target VNFs and modification of the packet; ii) dispatching the packet to the VNF; and iii) restoring the SR header after the packet is processed by the VNF.

VNFs can be divided into two categories based on their SR processing ability: 1) SR-aware; 2) SR-unaware.

*SR-aware network functions:* This type of VNFs are able to process an SR header based on which forwarding decisions can be made. Further, VNFs are also able to modify the SR header by pushing a new segment in the SR header.

*SR-unaware network functions:* This type of VNFs are not capable of understanding an SR header. They can only work based on a typical IP packet. In such a scenario, the SR/VNF connector forwards the IP packet to an SR unaware VNF by removing the SR header and re-applying the header once it receives the processed packet from the SR unaware VNF. In this case, adding the correct segment header back to the packet is important because the SR-unaware node may be part of multiple other service chains. Therefore, there is a need for a classifier to associate packets to the correct chain. Another possible solution to this problem is to associate a VNF to a single service chain.

ETSI proposed the NFV-MANO (NFV Management and Orchestration) framework for orchestration and management of NFV in provider networks [47]. Since adoption of NFV is an incremental process, therefore, MANO defines a scenario in which NFV-aware and NFV-unaware nodes co-exist [47]. The same approach can be directly mapped to the SR-aware and SR-unaware VNFs.

The authors in [48] proposed SERA (segment routing aware firewall) that is claimed as the first approach towards an SR-aware NFV applications. The authors proposed two modes of operation: In the base mode, IP tables are left intact (no modification), but the application is segment routing aware. The application accepts segment routed packets without the need of any proxy. In the advanced mode, the IP tables are now modified and the segment header specific actions are defined in the tables. The architecture uses SRv6 SFC [19] and the segment routing header is added to a packet. In the base mode, a preprocessor diverts the SR-aware and SR-unaware traffic. Thereafter, there is a unique path inside the firewall for both kinds of traffic. SR-traffic is analyzed in different blocks of the firewall architecture. The application handles SR packets in `encap` and `insert` modes. The original packet is extracted and firewall rules are applied. In the advanced mode, the SR specific actions are inserted in the IP tables. In this mode an extended action block is defined for performing SR-specific actions as shown in Table IV.

TABLE IV
EXAMPLE OF SR SPECIFIC ACTIONS

| Action | Significance |
|---|---|
| seg6-go-next | It sends the packet towards the next SID from SRH. |
| seg6-skip-next | It instructs SERA to skip the next SID of the SRH header. |
| seg6-go-last | It instructs SERA to skip the remaining part of segment list and process the last segment. |
| seg6-eval-args | The generic action that support SRH programmed actions. |

A special SR-specific actions block is defined known as *eval-args* to perform action based on SR-header values. Based on the current SID, SERA executes actions on a packet. Changes were made in the Linux kernel to incorporate the IP table modifications during the implementation phase. The results show performance degradation using SERA in comparison to basic IP table rules. The performance degrades when the number of rules increase. This is because of the design of IP tables that works in a stateless way and repeats all operations for each rule. Authors also provided an open-source implementation of SERA in [49].

### *D. Load Balancing (Segment Routing Load Balancer – SRLB)*

Load balancing can be classified under two categories: The first is network-level load balancing that uses ECMP and distributes load among paths. This approach does not take into consideration the state of the application. The second approach is application-aware load balancing in which load is distributed among applications by considering their states. For example, if an application is busy, a new query will be transferred to another idle application instance. This approach has a monitoring overhead, since the states can be acquired through monitoring only.

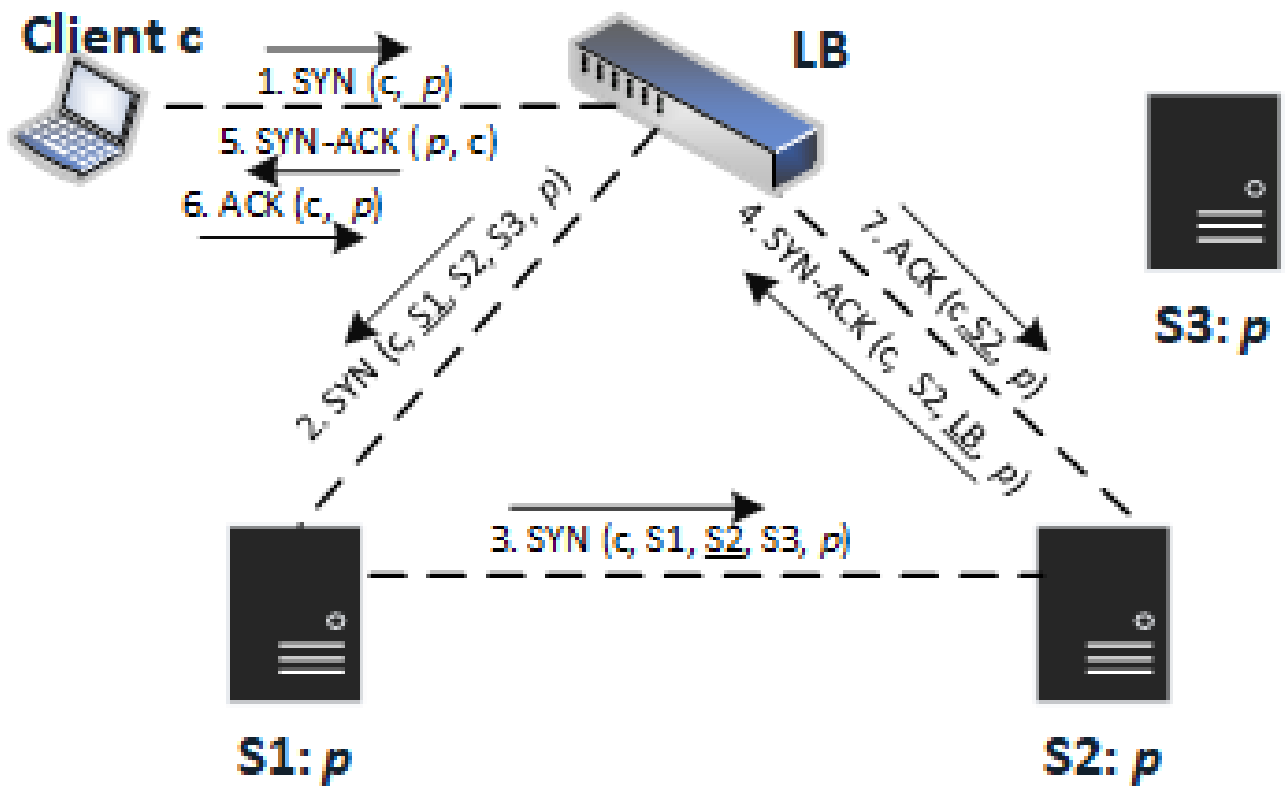

Fig. 11. An example to demonstrate SRLB.

An application-aware load-balancing scheme with the use of segment routing is presented in [50] that takes into consideration the state of application without any monitoring overhead. *Service hunting* is defined as a process for service selection in the network using segment routing. For example, consider a scenario in which an application is hosted at different servers. The request arrives at a load balancer already aware of the servers running the application instances. From the candidate servers, the load balancer selects a set of servers and appends them to the request through the segment routing header. A segment in the header corresponds to a server. Once the first server receives the request, it checks with a virtual router installed at the server for the availability of resources in that server. The virtual router is assumed to be aware about the application status (load on the application). If the server is willing to accept the request, then it signals to the load balancer through a segment-routed header inserted in the TCP SYN-ACK packet. The flow corresponding to the request is pinned to the server after the acceptance from the server. If the server is not willing to accept the request, then the request is transferred to the next segment in the header. This process repeats till all the servers in the segment list are traversed. Shown in Fig. 11 is the working of the SRLB scheme. A client requests the load balancer present on the edge of a data-center for an application *p*. The load balancer has pertinent information of whether *p* is running on servers *S1*, *S2* and *S3*. Therefore, the load balancer adds a SR header consisting of the addresses for *S1*, *S2* and *S3*. A request packet is sent to the server *S1* that refuses the request. Now the packet is forwarded towards the next server in the SR header, i.e. server *S2*. *S2* accepts the request and the client is notified about the acceptance. A connection between the client and server *S2* is established.

To guarantee service, the penultimate segment can indicate to the application that it cannot rejects a query. Random selection and consistent hashing are some of the approaches that can be used as a selection method for servers from the candidate servers to be included in the segment list of the header. The results in [50] indicate that SRLB is able to better spread the load among all the servers than a random load-balancer. It was shown in [50] that the mean response time of a SR based LB is up to 2.3x better than a random LB scheme. Moreover, SRLB better spreads queries between the servers and achieves fairness index closer to 1.

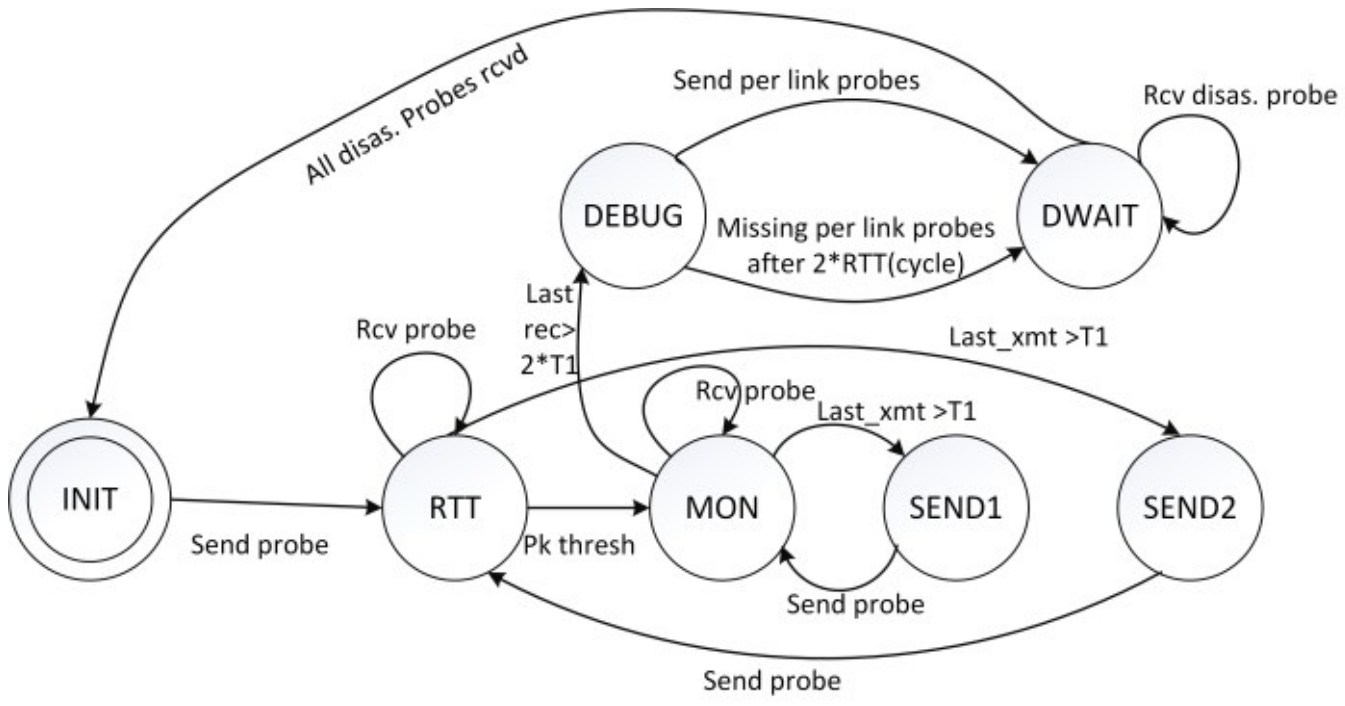

Fig. 12. Service monitoring state diagram.

### E. Service Monitoring

Traditionally, probing [51] is used to monitor a network. The problem with traditional probes is that they can never cover the unused links. If the probes were to be sent over the tunnels formed due to RSVP-TE, then the resulting signaling overhead would be significant. In [52], authors used the segment routing architecture for monitoring a network, where a model called SCMon sends and receive probes over cycles that are created in a way that they cover all the links of the network. The cycle computing algorithm takes an input parameter $k$ along with the graph of a network that specifies the maximum number of segments that can be used for a cycle. The segments generated by the algorithm for a cycle are ECMP-free. This means segments are associated with only one segment list. The segment routed paths are created for each cycle in the network. The work is of importance as it is the first to demonstrate segment routing for monitoring the network.

Fig. 12 shows the state diagram used for the operations of SCMon. Initially, SCMon enters into an *INIT* state that calibrates the *RTT* for each cycle. SCMon then starts to send probes over the cycle based on a predefined rate, *T1* (that was calculated experimentally to be 0.2 ms). The other parameter $P_k$ (defined as the number of configurable probes) is also set to a value that is large enough to account for unexpected jitter. Once SCMon receives $P_k$ probes, it assumes the cycle to be up and enters in the *MON* state. If a probe is not received in a duration of *2T1*, then the cycle is considered to be timed-out and SCMon enters into a *DEBUG* state. In this state, it sends one probe per segment to determine the faulty segment. If all probes were received within a duration of $2RTT_{cycle}$, then the cycle is considered as a backup and SCMon re-enters the initial state. This helps to recalibrate *RTT*. The authors in [46] have shown blackhole detection using the SCMon in different topologies. Most of the blackholes are detected within less than 100 ms using SCMon.

### F. Traffic Duplication

Traffic is duplicated in a network for providing 1+1 protection on the end-to-end paths for latency sensitive TCP applications and also for the purpose of monitoring by using port mirroring. Multicast also requires traffic duplication. A segment routing-based traffic duplication using the Linux kernel has been implemented in [53] for providing 1+1 network protection. An algorithm is proposed to identify segment-able disjoint paths on which data can be duplicated. The algorithm limits the number of segments that a header can carry. Therefore, a shortest path is divided into a $k$-segmented path to reduce the number of segments in the header.

### G. Multidomain networking using segment routing

A network may contain multiple domains where each domain is separately controlled by an SDN controller or PCE. A domain can be based on the topological distribution, policy or rules and can be created for reducing the complexity and load on a single controller. In such a scenario, the controller or PCE of different domains needs to share measurements of their respective domains with an orchestrator in order to provide for an end-to-end service. This might require the deployment of multiple LSPs for monitoring, and can lead to scalability issues. Additionally, protocols for the controller to orchestrator communication do not provide any specific handling and advertisements of the collected information. In [54], the extension for BGP-LS protocol was proposed as a method to encompass retrieved information within the controller to the orchestrator. BGP-LS is an extension to Border Gateway Protocol (BGP) for distributing the network's link-state (LS) topology model to external entities, such as the SDN controller. The work in [54] considers mainly the network delay and defines two types of probes for monitoring. The first probe is originated by the node e.g. MPLS ping or Bidirectional Forwarding Detection (BFD) messages. The second probe relies on external monitoring systems that facilitate in injecting probes at different locations. Minimum and maximum delay values are calculated for paths within the domain. A controller advertises the intra-domain links between the border router and the orchestrator. In [54], BGP-LS is modified to include the computed max-min delays in the advertisements between border nodes within a domain. Hierarchical PCE is used to implement an orchestrator for controlling inter-domain controllers. Implementation of this use-case is discussed in the next section.

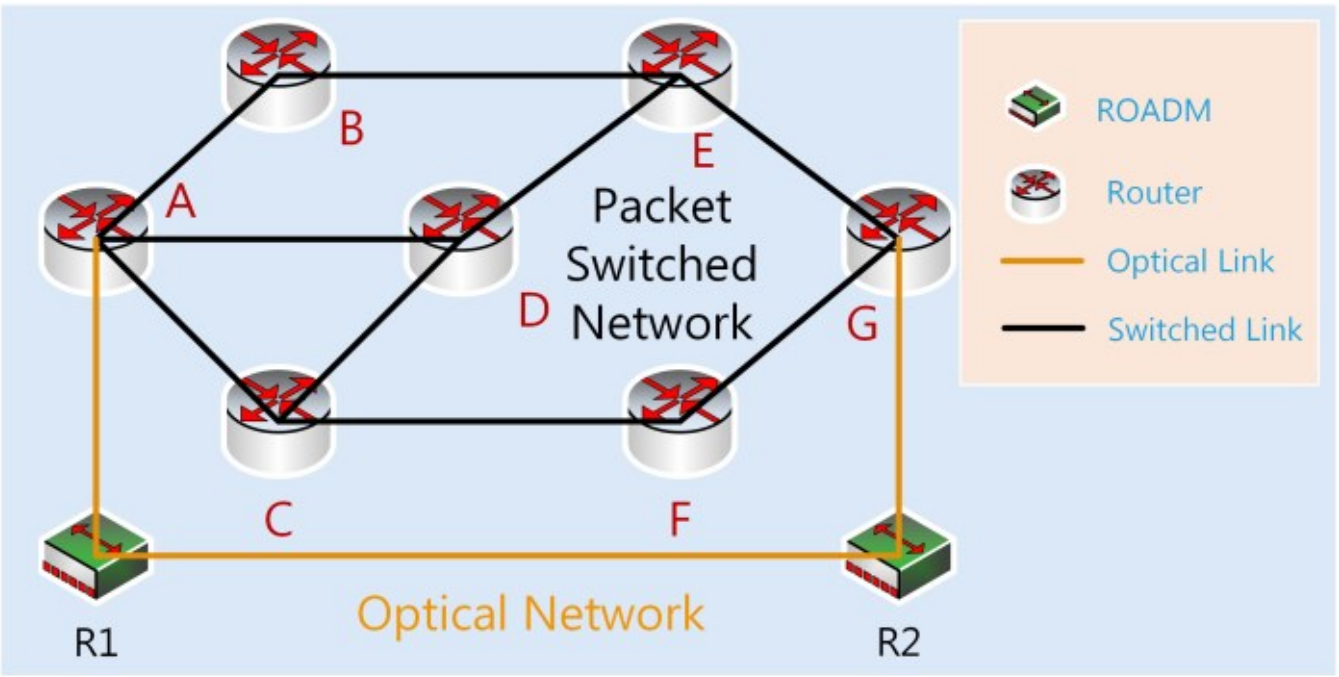

Fig. 2. Multi-layer network.

### H. Segment Routing in Multi-layer Network: Optical bypass

Segment routing can be effectively utilized to route a packet in a multi-layer network. In [55], authors have demonstrated the use of segment routing for achieving optical bypass. A centralized SDN controller is used to compute the path based on preset policies. The controller generates a segment list. This generated list is configured at the node for packet forwarding. The optical bypass is demonstrated in a packet-switched network in Fig. 13. In this figure, we observe that Reconfigurable Optical Add Drop Multiplexers (ROADMs) are available at some of the nodes. These ROADMs are connected using 10Gb/s/OTN muxponders (that multiplex several slower speed signals into a higher speed signal). The SDN controller is used to compute the segment-routed path based on the requested bandwidth. In case the bandwidth request exceeds a certain threshold, then the controller provides a segment-routed path using the optical bypass. If however, the bandwidth request does not exceed a threshold, the controller provides the segment-routed path in packet-switched network. For instance, in Fig. 13, whenever there is a high-bandwidth service request between node-*A* and node-*G*, the request is routed over the optical network through ROADM *R1* and *R2*. For low bandwidth requests, forwarding takes place only in the packet-switched network through nodes *C* and *F* (multiple paths can be possible).

## V. IMPLEMENTATION

In this section, we discuss the various implementation schemes associated with segment routing that are presented in literature.

In [56], two SR implementations are presented. The first implementation focusses on SDN-based segment routing which is further applied to an optical bypass use-case [55]. Packet switch capable (PSC) nodes supporting an IP/MPLS data plane are considered in this deployment. A PSC node in its routing table includes a set of MPLS based rules. The SIDs are mapped to MPLS labels. The SDN controller assigns these SIDs to all the nodes in the network. The SR controller manages the network topology. The SR controller computes paths and generates a list of segments. The controller then configures the ingress node for each path by this generated list of segments to modify a packet and adds the SR header to the packet. To support an MPLS data-plane, OpenFlow 1.3 [15] was used for communication between the controller and the PSC nodes. The architecture of the SDN controller is shown in Fig. 14. The dashed-line components are the new implementation modules while the other modules shown in the figure are traditional components used by the SDN controller [57].

The new modules added to SDN controller architecture are *request handler*, *network tracker*, *SR engine* and *per-flow monitor*. The *request handler* is a north bound interface that receives connection requests from applications. The *network tracker* keeps track of a network node and associated interconnections. The *per-flow monitor* gathers the statistics of the flows that are configured in the network. The *SR engine* runs the SR algorithms for the selection of the path and the generation of the segment list for packet forwarding.

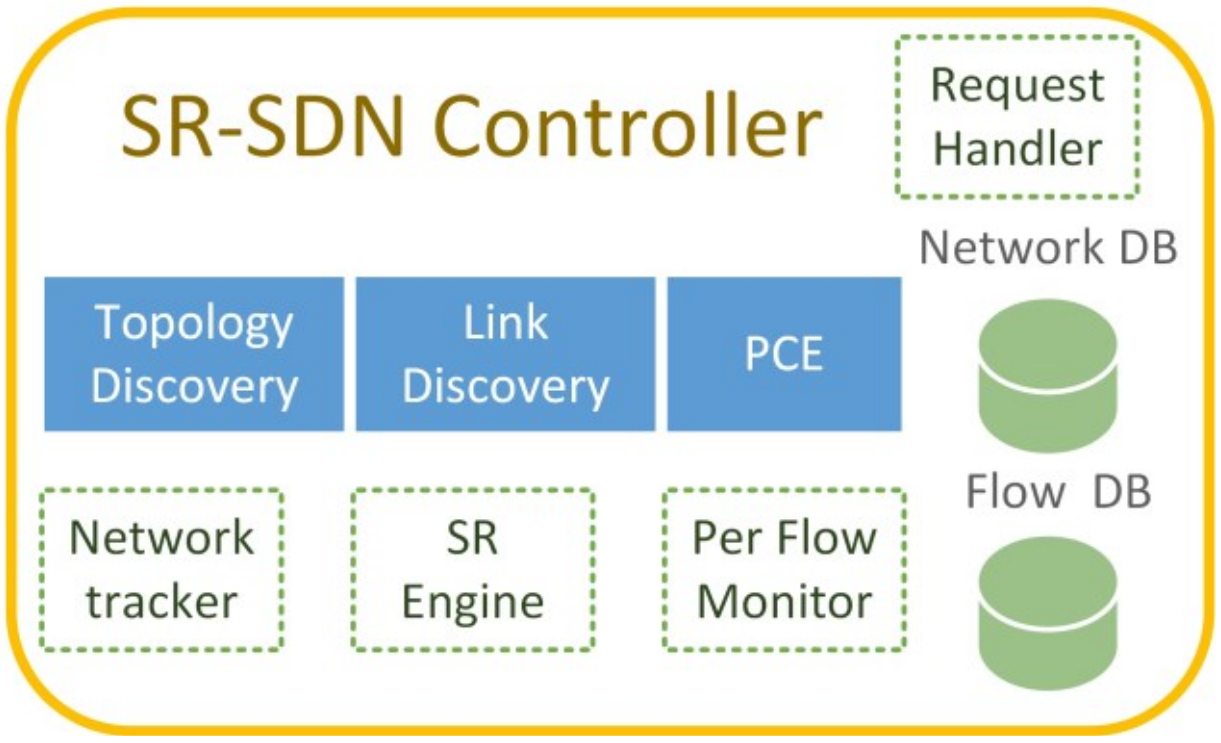

Fig. 3. SDN SR controller architecture used in [56].

The second implementation in [56] is an extended version of PCE supporting segment routing and is used to perform connection request handling and path selection. In this implementation, a node behaves like a path computation client (PCC) and the SDN controller functions as the PCE. The SDN controller stores all the states of the nodes present in network in its database. The database is updated based on the PCEP notification messages that are received from the nodes. Each router is equipped with an external software agent. This agent operates through the northbound of the PCC and establishes a connection with the PCE. On the southbound interface, the software agent collects IGP based information and other shortest-path related information from the node. The data-plane is implemented using commercial IP/MPLS routers. The SIDs are obtained by the combination of router ID and a constant prefix of 10000 to exploit static MPLS range of labels. Authors claimed that both the implementations were successfully utilized to demonstrate dynamic flow re-routing and no packet loss was experienced during rerouting.

Different approaches are proposed for multidomain SR and presented in [58]. Network control planes of different domains can interact with each other either directly or through a multi-domain orchestrator. In one of the approaches, end-to-end segment routing is considered. A segment list is added to a packet at an ingress node in the path and no modification is made by any of the nodes in the path till the packet reaches its destination. In another approach, domain-specific segment routing is considered. A segment list is added in the packet at each ingress node of the SR-domain traversed by the packet. In this approach controllers of the domains do not directly interact with each other.

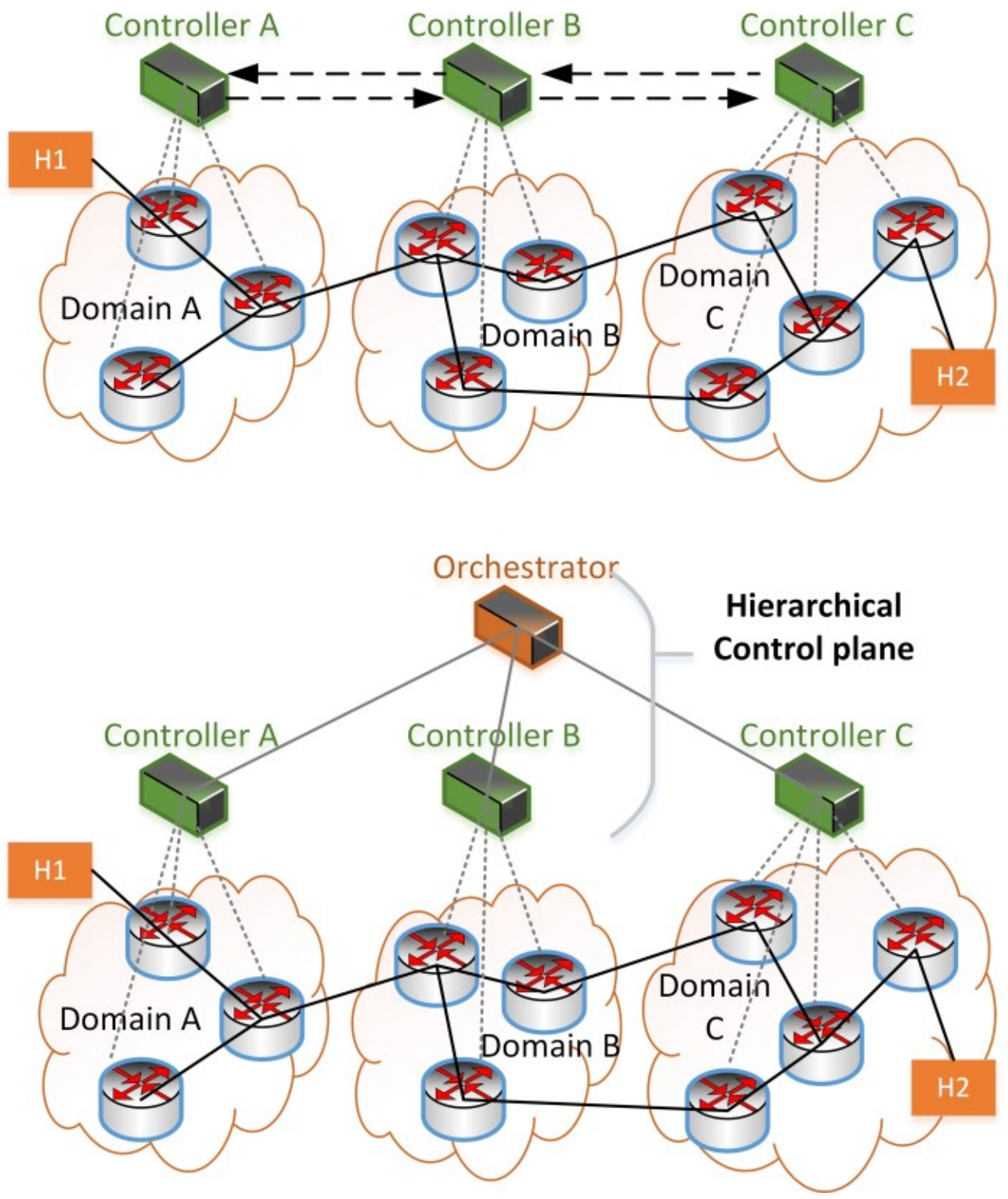

Fig. 15. Multidomain network with controller talking to each other (Top) and hierarchy of controllers (Bottom).

In the end-to-end approach, the controller of the destination node sends the segment routed path information of its domain to an upstream controller. This upstream controller appends the segment routed path information of its domain and sends the path information further towards its upstream. The process is repeated until the complete path information reaches the controller connected to the ingress node. With this process, the ingress node's controller gets the complete path information pertaining to the segment routed path. Using this information, the ingress node's controller generates the segment list for packets to be forwarded. Shown in Fig. 15 (Top) is an example of a multidomain network where controllers interact with each other. Let us consider a packet that needs to be forwarded from host H1 to H2. Therefore, the controller-C, which is connected to the egress node computes the SR path of domain-C and forwards this path information to controller-B. Controller-B then computes the SR path in its domain and appends the computed SR path to the path information that it has received from controller-C. Thereafter, controller-B sends this path information to controller-A. Controller-A manages the ingress node in domain-A and computes the SR path that a packet must follow in domain-A. Thereafter, controller-A combines the SR path computed for domain-A with the path information received from controller-B to generate the segment list. This list is then used for forwarding the packet along the computed path.

In the second approach, a hierarchical control plane is presented for segment routing. The network advertises the SIDs in its own domain. This implies that the controllers of other domains do not have any information pertaining to the intra-domain topology under the premise of other controllers. An orchestrator exists that resides on top of all the controllers and interacts with each of them using a Northbound RESTful API (of the controllers). IS-IS protocol is used to exchange route information and SIDs between the routers within a domain. Standard southbound interfaces such as BGP-LS and PCEP of the SDN controller are used to obtain the network topology. An example of a multidomain network with a controller is shown in Fig. 15 (Bottom). The experiments with multi domain MPLS network using OpenDayLight (ODL) are presented in [58]. ODL is configured to communicate with BGP-LS routers in each domain so as to get topology information. Paths in a network can be created using a prefix SID or an adjacent SID. For prefix SIDs, routers use the shortest path provided by IGP. For adjacent SIDs, routers forward packets directly to the specified interface. The results in the paper show that the hierarchical control-plane approach outperforms the end-to-end approach in terms of the maximum segment list depth.

*IP to SR transition:* Transition from a pure IP network to a segment routed network can be critical and difficult to manage for a service provider. An incremental deployment scheme to support traffic engineering is presented in [59]. This work reveals the possibility of incremental upgradation of the network with considerable improvement of the network performance even with a reduced number of SR nodes, whereby instead of fully migrating from an IP to an SR network only a few select nodes are replaced as an SR node resulting in a hybrid IP/SR network.

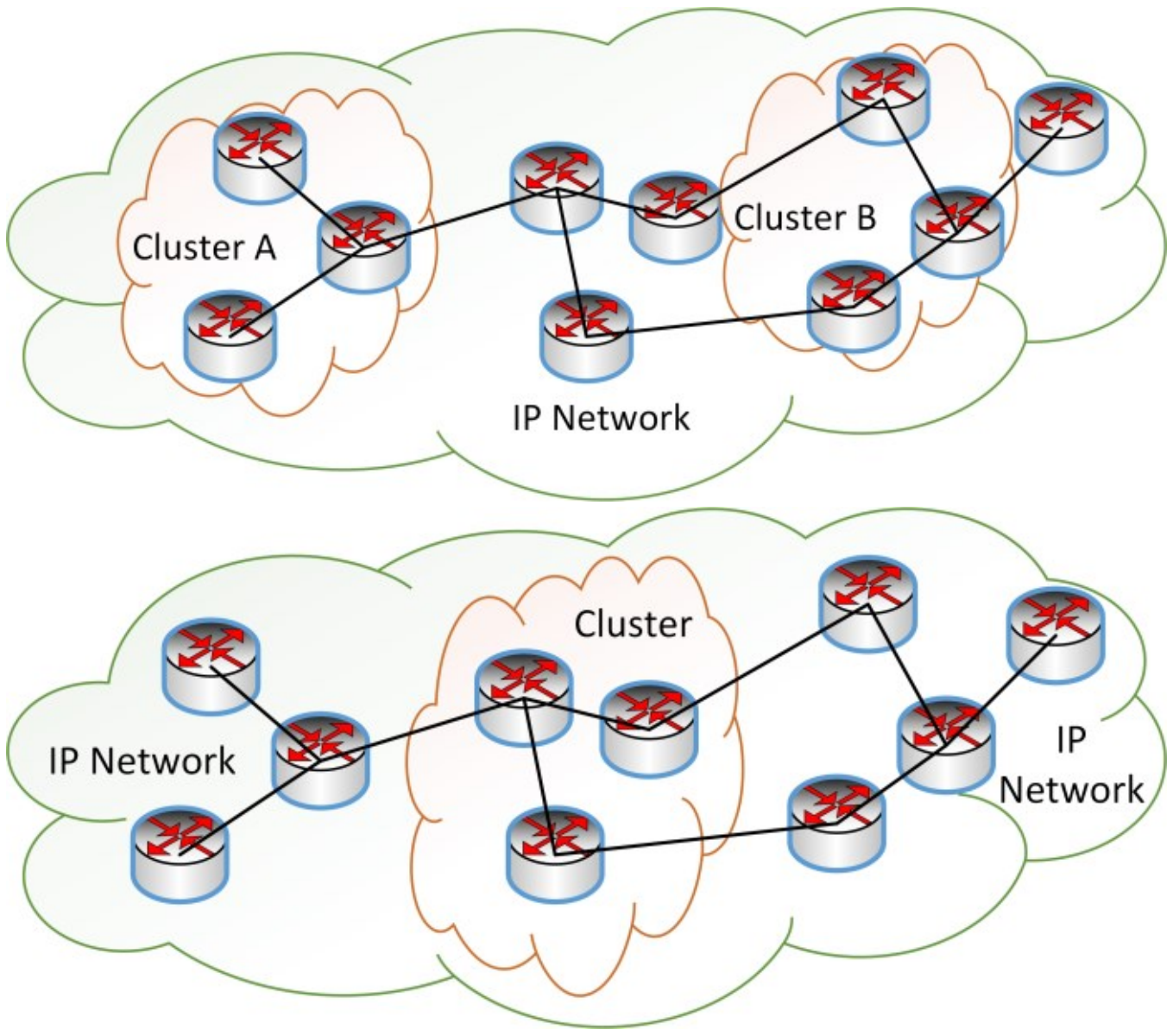

Fig. 4. Example of single SR cluster (Top) and multiple cluster (Bottom).

Performance evaluation in [59] shows that a hybrid network is comparable to a fully SR network. The deployment of SR nodes in the network was done in a manner such that they form a single or multiple cluster as shown in Fig. 16. In the figure, a cluster represents an island of SR nodes. A problem formulation based on MILP is used to identify the set of nodes to be replaced by an SR node. A cluster is bound to a maximum number of SR nodes which it can have. The goal of the optimization is to minimize the maximum link utilization with the assumption that the traffic matrix representing a forecast for a sizable time-period is available. The output of the optimization problem is the set of SR nodes forming a cluster and the routing of paths for the flows transiting through the SR cluster. Two traffic scenarios are considered based on packet delivery: (1) order aware and (2) order unaware. In the case of order aware, all the packets of a flow follow the same path and packets are delivered at the destination in the order in which they arrived at the ingress. For order unaware, the traffic of a flow can take different paths based on the path utilization. As a result, there is no guarantee on the packet order at the destination. Three different set of analysis are performed for the proposed deployment. The analysis evaluates the effectiveness and stability of the incremental deployment.

- *Potentiality of the proposed SR-Domain (SRD) approach*: This analysis is performed to characterize the TE performance. The authors consider maximum link utilization as a key parameter to evaluate performance. In shortest-path routing, a link gets congested when a large number of paths are mapped through the link. As a result, no more traffic can be accommodated for those source-destination pairs that have paths over such links, even when sufficient bandwidth is available on the other non-shortest paths. Due to this behavior of shortest path routing, maximum link utilization increases in the network. Traffic engineered paths have the capability to reduce the maximum link utilization by utilizing the available non-shortest paths. Therefore, maximum link utilization becomes an important parameter while evaluating the TE performance. It is observed by the authors in [59] that as the network moves from a pure IP to a SR network the maximum link utilization decreases progressively. This indicates the effective utilization of links by the traffic engineering paths. Another set of results by authors in [59] lead to a conclusion that as compared to a single cluster, multiple clusters provide better performance. However, when the cluster size is significantly large the performance of a single cluster is almost similar to the multiple cluster scenario. This analysis suggests that for a cluster of small size, it is desirable to have multiple clusters in the network, whereas for a large sized cluster, it is enough to have single cluster for optimized performance. This behavior is counter-intuitive due to the fact that there is more opportunity of traffic engineering in a large cluster as compared to a small cluster. As a result, the performance of a large cluster is comparable to multiple smaller clusters.

- Stability analysis: This analysis is carried to determine the robustness of the optimization problem with respect to the variations in traffic. Although a forecasted traffic matrix is considered in the problem formulation, yet the actual traffic matrix can be significantly different. For experiments, a noise matrix has been added to the traffic matrix. This analysis is performed for different values of cluster size, number of clusters and variations in traffic. Results proved that Irrespective of traffic variations, the relative gap in maximum utilization is independent of noise between the hybrid and SR networks.
- Topological analysis: This analysis is performed to determine the topological parameters that can help to identify the IP node that can be replaced by an SR node. Three parameters are considered in this analysis: 1) degree of a node; 2) node betweenness and 3) traffic (most loaded links). It is observed that first two parameters do not help in replacing the node with SR nodes in an IP network to obtain an optimized hybrid network (MILP solution). On the contrary, the most loaded link parameter showed strong correlation with the SR node present in the optimized network. This correlation is due to the ability of a SR node to utilize the traffic engineered paths to minimize the maximum link utilization.

## VI. Segment Routing in Commercial Equipment

Segment routing feature has been added by multiple vendors such as Cisco, Juniper, Arista, Alcatel-Lucent etc. in their routing platforms.

Cisco ASR 9000 Series Aggregation Services Routers [60] can be configured for segment routing using Cisco IOS. Segment routing in Cisco platforms rely on a small number of extensions to IS-IS and OSPF protocols. Cisco restricts the use of segment routing in ASR 9000 to only IPv4 and IPv6 data-plane when the IS-IS mode is enabled.

Juniper supports segment routing in its MX series routing platform, which can be configured using the JUNOS OS [61]. The Junos OS uses BGP-LS peering or IGP adjacency to compute the network topology. For segment routing feature to be included, Junos OS does not support OSPF.

Arista uses its Extensible Operating System (EOS) and R-series platforms [62] with FlexRoute capabilities to support segment routing. This solution from Arista provides segment routing using the MPLS data-plane. Arista also claims that R-series platforms does not have any restriction on the size of the MPLS label stack.

Alcatel-Lucent (now Nokia) supports segment routing through its operating system SR OS [63]. SR OS supports both IS-IS and OSPF protocols.

Barefoot networks have also demonstrated segment routing support for IPv6 data-plane using their P4 programmable Tofino switch [16].

Huawei NE40E [64] also uses segment routing with OSPF and IS-IS IGPs for distributing SIDs. Using segment routing they claim to achieve multiple benefits such as: achieving topology independent-loop-free alternate (TI-LFA) protection, a simplified MPLS control plane and better evolution to SDN networks.

Ericsson 6000 series routing platform [65] also included segment routing functionality. The segment routing utilizes the SDN control plane.

Ciena also included segment routing in their Route Explorer application that is part of Blue Planet Route Optimization and Assurance products [66]. SR with an SDN application can be used to provision TE tunnels automatically. It can also be used to provide value-added services such as bandwidth calendaring, bandwidth management, and bandwidth on-demand.

TABLE V
Comparison of the different products supporting segment routing

| Product and Spec | Cisco ASR 9000 [60] | Barefoot [16] | Juniper MX series [67] | Arista's R-series [62], [68] | Alcatel-Lucent [63] | Huawei NE40E [64] | Ericsson Router 6000 [65] |
|---|---|---|---|---|---|---|---|
| Capacity | Up to 160 Tbps | 6.4 Tbps | Up to 80 Tbps | Up to 12 Tbps | Up to 9.6 Tbps | Up to 81.92 Tbps | Upto 2.1 Tbps |
| Line rates (Gbps) | 100 | 10, 25, 40, 50, 100 | 1, 10, 40, 100 | 10, 25, 40, 50, 100 | 10, 40,100 | 10, 40, 100 | 10, 100 |
| Segment routing compliance? | Yes | Yes | Yes | Yes | Yes | Yes | Yes |
| Dataplane supported | IPv4 and IPv6 (IS-IS mode) | Programmable to support any data plane | IPv4 and MPLS | MPLS | MPLS | MPLS, IPv4, IPv6 | IPv4, IPv6, MPLS |
| Latency | - | - | - | 4 μs | - | - | - |

A comparison of some of the different routing platforms supporting segment routing is presented in Table V. It should be noted that Table V is not an exhaustive list as several vendors have published only limited information pertaining to SR implementations due to the nascency of the field.

## VII. Segment Routing Positioning and Interoperability

We now describe the positioning and deployment of segment routing in various scenarios. We also describe interoperability of segment routing with other protocols.

### 1) *Segment routing in Service provider network*

A service provider network is a large network potentially providing connectivity across a large region. A service provider network is divided into sub-networks such as edge or access, Metropolitan (metro) and core networks. Edge networks provide last-mile connectivity to the end-users by connecting them to central offices. Metro networks provide city-wide connectivity by connecting central offices located across a city. Core networks are backbone networks and usually connect cities with each other. Core networks also provide peering points to interface with other service provider networks. Examples of such service provider network are the AT&T and Verizon networks, which are spread across the country and connect end users. In such a large network, it is crucial to manage and provision services in a manner that efficiently utilizes the network resources and also adhere to service level agreements. Therefore, service provider networks can be benefited by segment routing in terms of traffic engineering, which leads to effective utilization of network resources. Other than traffic engineering, a provider network can also take advantage of segment routing for network restoration in the event of a fault/failure in the network. In large provider domains, segment routing can be used to connect across islands of connectivity – core networks can use segment routing in the backbone to connect large P-routers to each other. This will result in cost savings as well as better efficiency across the domain due to the virtue of lower processing required at nodes across the network.

### 2) *Segment routing in a large enterprise network*

Large enterprise networks comprise of routing equipment, firewalls, intrusion detection systems (IDS), intrusion prevention systems (IPS) and servers to host enterprise applications. An enterprise network connects geographically isolated offices, users and workgroups. An example of such a network is a banking network where branches of a bank in different parts of a city/country are connected to each other.

Segment routing can be used in an enterprise network to provide load-balancing as well as to provide data access to servers/data-centers, in addition to providing traffic engineering for routing. Segment routing can be a simpler way of provisioning L2VPNs and L3VPNs making the solution cost-effective and scalable. Segment routing can also be utilized for monitoring network statistics.

### 3) *Segment routing in a campus network*

The campus network is an example of small enterprise typically limited to a single geographic location/office. For example, consider a university network such as Stanford University campus network that provides connectivity to its students, faculty, administration and various other entities on campus. Due to the small size of the network the opportunity for traffic engineering is limited. Despite the size limitation segment routing can be used in such type of a networks to provide the fast reroute functionality in an event of link/node failure. A campus network can also make the use of segment routing for network slicing – giving higher quality access to a select subset of users. Network slicing is reported in detail in [69]

### 4) *Segment routing in datacenter network*

A data center is an important part of a communication network, which provides resources for storage, computation and networking. Most over-the-top operators have large data-centers such as Facebook, Google etc. A data center network plays a critical role in providing interconnection between its constituent resources. Segment routing can be potentially used in a data center network to provide load balancing among the servers hosting same application/content. The lower overhead of segment routing as well as source-routing capability is ideal for implementation in large data-centers.

### *SR Positioning: label distribution comparison*

SR label distribution using IGP is similar to MPLS label distribution using LDP. In both schemes, labels are automatically distributed when an adjacency peering is formed. Both LDP and SR form stateless tunnels. Stateless tunnels are end-to-end paths that do not require any signaling protocol for the creation of tunnels. However, LDP requires that IGP should be used and synchronization between both the protocols (LDP and IGP) is needed to reduce traffic blackholing [70]. Traffic blackholing is when packets are dropped en route due to insufficient capacity or improper configuration at intermediate routers. Further, labels in LDP are of local significance as opposed to SR where labels can have local/global significance.

RSVP-TE is widely deployed for fast reroute and traffic engineering capabilities. RSVP-TE networks are built using a full-mesh of point-to-point tunnels with a requirement to configure the state of the control plane and data plane at each of the nodes along a path. In SR, only ingress nodes are configured with a traffic-engineered path. With an SDN controller, SR can facilitate a fine control over traffic engineered path as any number of the constraints can be used to define a path which is limited in case of RSVP-TE.

## VIII. Conclusion

We have provided a comprehensive survey on segment routing. Segment routing can be seen as a crucial improvement over MPLS in terms of traffic engineering and can prove as an important step towards the assimilation of SDNs for telecom service providers, as well as enterprise and campus networks. Due to the minimalistic nature of enhancements required, SR can be implemented only through control plane manifestations (which we have shown as examples from commercial vendors). This makes SR an excellent SDN migration technology for providers. In this survey paper, we have initially discussed the need for SR, its positioning among other technologies such as MPLS, Carrier Ethernet and SDN. We have also considered engineering aspects of SR, including key optimization issues. Our focus has been on label encoding techniques that are important because of the ability of SR to seamlessly integrate with existing technologies. Thereafter we consider SR use cases in contemporary networks such as protection, forwarding, optimization, as well as in future provider implementations such as SDN and NFV. We then focus on implementation in lab environments and results from test-beds, gearing towards industry assimilation of segment routing that is summarized as a comparison across product families. In summary, segment routing is a promising technology that can be implemented across the spectrum – in campus domains, enterprises, data-centers and of course the most important use case of large provider networks. Segment routing is a smarter way of using the MPLS forwarding plane and an efficient approach towards SDN adaptation. The carrier-class aspects of segment routing imply that it would be an important technology for providers to consider in the future.

# APPENDIX

## 1) Control Protocols and their SR extensions

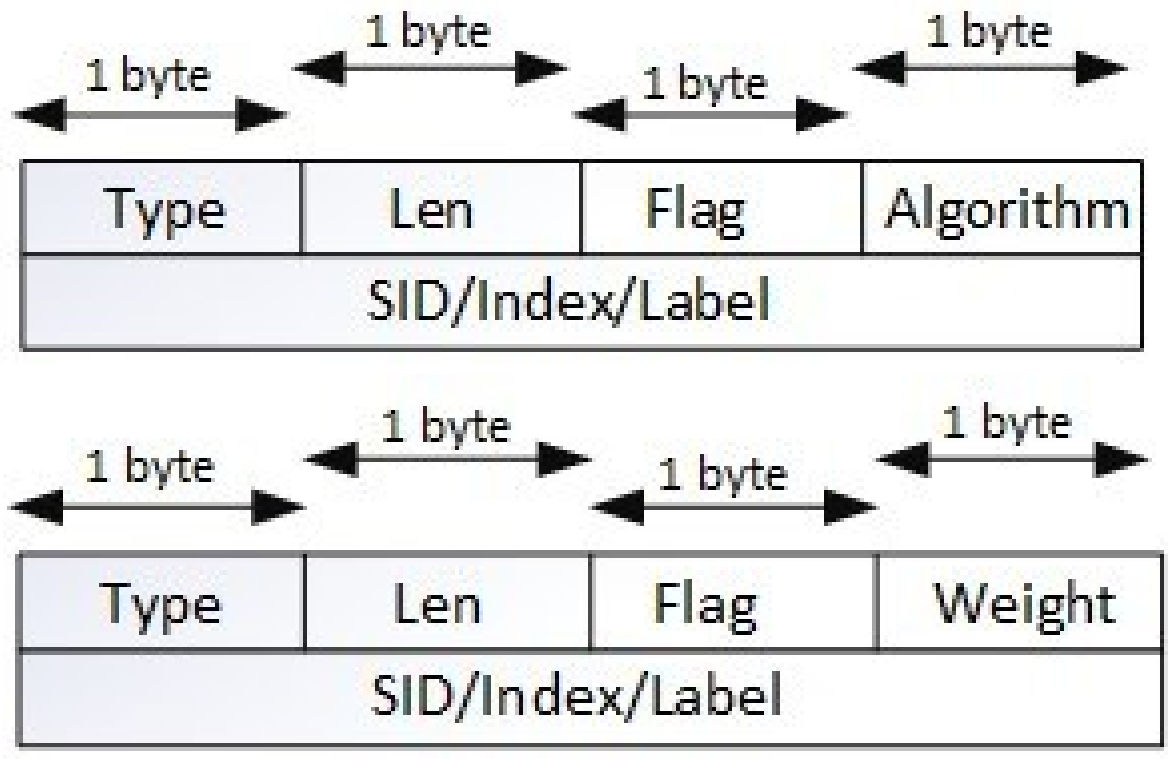

Fig. 17. ISIS sub-TLV format. Prefix SID (Top) and Adjacency SID (Bottom).

### *a) Intermediate Systems-Intermediate Systems (IS-IS) and SR*

IS-IS is a link-state routing protocol widely used in the provider networks. The IS-IS protocol has been extended to support segment routing [24]. We describe some salient features in this extension. IS-IS provides encoding for the SIDs, adjacency SID, LAN-adjacency SID and binding SID. We now describe the prefix SID and adjacency SID in following subsection.

- The Prefix SID represents ECMP-aware shortest path to a prefix. Shown in Fig. 17 (Top) is the format for the prefix SID sub-TLV advertised by a router in the domain. A prefix SID must be unique in an IGP domain (though this is relaxed when the *L* flag is set in the TLV – type length value fields). A prefix SID must carry the index of an actual SID (when the *V* flag is set). The index is used to determine actual SID/Label value from the set of SID values advertized by the router. The router advertizes a prefix SID for any prefix reachable from the router. *Type* value is undecided and *Len* is a variable. The Flag field contains 6 binary flags, which are: 1) *R* Flag, which is a re-advertisement flag. If the *R* flag is set, then the prefix to which this prefix-SID is attached is sent from the router; 2) *N* flag, which when set denotes that the prefix-SID that represents the node SID; 3) The *P* Flag, which when set indicates that the penultimate node must not pop the prefix SID from the packet before distributing the packet to the node that advertized the prefix SID; 4) The *E* Flag, which when set indicates that any of the upstream router from the originator must replace the prefix SID with a null value; 5) The *V* Flag, which when set indicates that Prefix SID must carry a value instead of an index; 6) Finally the *L* flag, when set, determines that the value carried by the prefix SID is of the local significance.

Apart from the 6-bit flags, the other two bits of the flag are ignored. The algorithm field specifies the used algorithm for calculating the reachability of the prefixes attached to the prefix SID. Example of such algorithms include shortest path first (SPF), constrained SPF, strict SPF, etc. Based on the values in the *V* and *L* flag, SID/Index/Label tag contains either a 32-bit index defining the offset in the SID/Label space advertized by this router or a 24-bit label where the 20 rightmost bits are used for encoding the label value.

- The adjacency SID represents a hop over a specific adjacency between any two nodes in the IGP. The adjacency SID is local to an advertising node. The format of Adj SID sub-TLV is shown in Fig. 17 (Bottom). The format of Adjacency SID sub-TLV is similar to the prefix-SID, except for the changes in flag field and an addition of weight field. The Adjacency SID sub TLV has 6 flags. These are: 1) *F* flag: If the *F* Flag is set then the Adj SID refers to IPV6 encapsulation, otherwise the Adj SID refers to an IPv4 encapsulation; 2) *B* flag: If the *B* flag is set then the adjacency SID is eligible for protection; 3) The *V* and *L* flags: The *V* and *L* flags are similar to prefix SID sub-TLV; 4) The *S* flag, if set implies that Adj SID represents a set of adjacencies; 5) The *P* flag is called the persistent flag, which if set, represents that Adj-SID is persistently allocated and remains the same even after restarting an SR capable router or interface. The *Weight* field in Adj-SID is used for the load balancing.

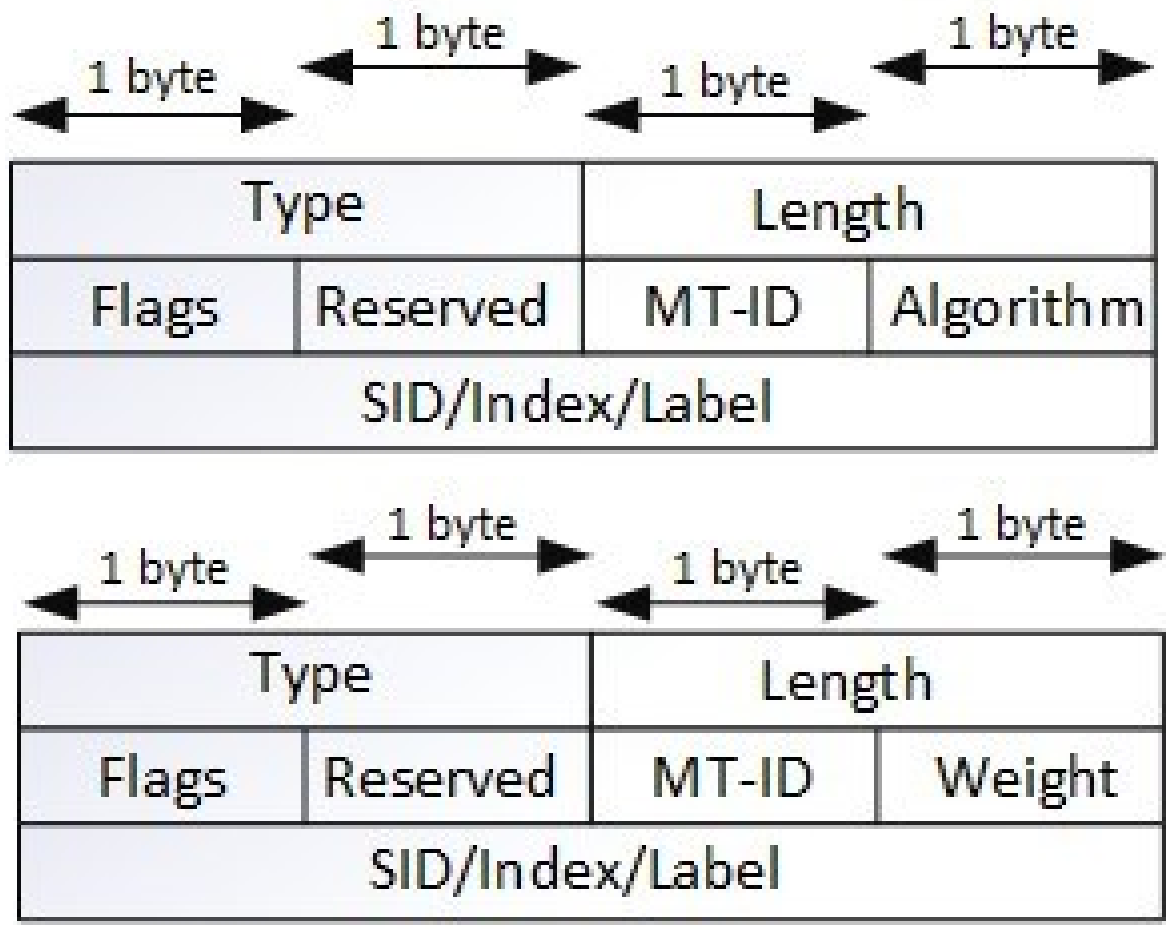

Fig. 18. OSPF sub-TLV format. Prefix SID (Top) and Adjacency SID (Bottom).

### *b) Open Shortest Path First (OSPF)*

When using OSPF, for propagating the SID, the prefix-SID sub-TLV and adjacency-SID sub-TLV are defined (similar to IS-IS) [25]. We explain the exchange format shown in Fig. 18 for these two methods:

- Prefix SID Sub-TLV: In this format the *Type* field is not standardized and length is a variable. *MT-ID* is defined as Multi-Topology ID. OSPF uses Multi-topology ID to compute multiple topologies and find paths to IP prefixes of each MT independently. The flags field has 5 flags of 1 bit each. These flags are as follows: 1) *NP-Flag*: If this flag is set, then the penultimate router must not remove the Prefix-SID before delivering the packet to the node that advertized the Prefix-SID; 2) *M-Flag* is used for mapping to a Server Flag. If set, the SID is advertised from the Segment Routing Mapping Server functionality; 3) The functionality of *E-Flag*, *V-Flag* and *L-Flag* and is same as the IS-IS prefix SID sub-TLV. The algorithm and SID/Index/label fields are also similar to the IS-IS prefix SID sub-TLV.
- Adjacency-SID sub TLV: Adjacency SID in segment routing is similar to IS-IS except for the flag values and weight field. In the adjacency-SID sub TLV there are 4 flags defined in sub-TLV and these are *B*, *V*, *L* and *S*. The *B* Flag is the backup flag and when set to 1, the adj-SID is eligible for protection. The *V* Flag and *L* Flag is the same as the prefix-SID sub TLV. The *S* Flag, if set, indicates that adjacency SID represents a set of adjacencies. *Weight* is used for the purpose of load balancing.

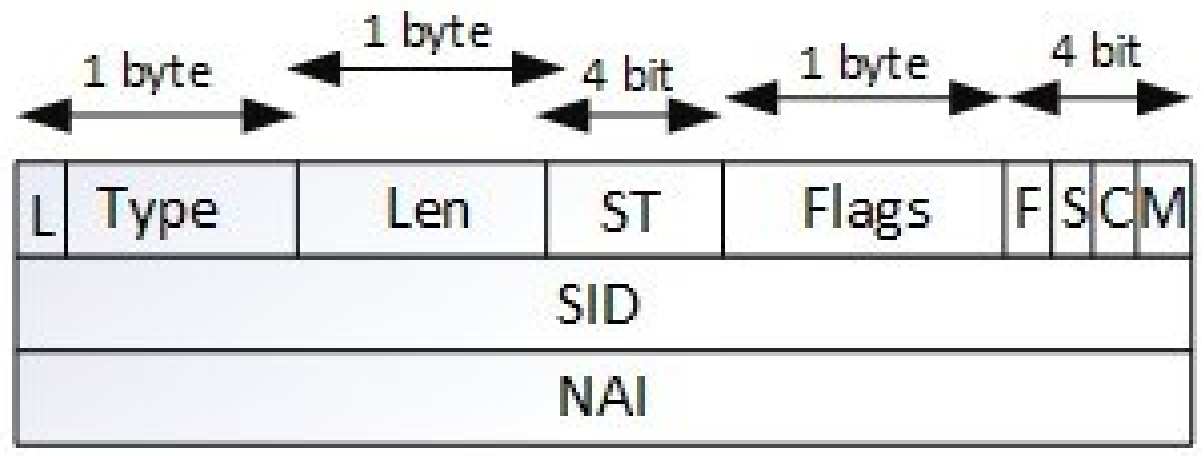

Fig. 19. SR-ERO Sub-object format.

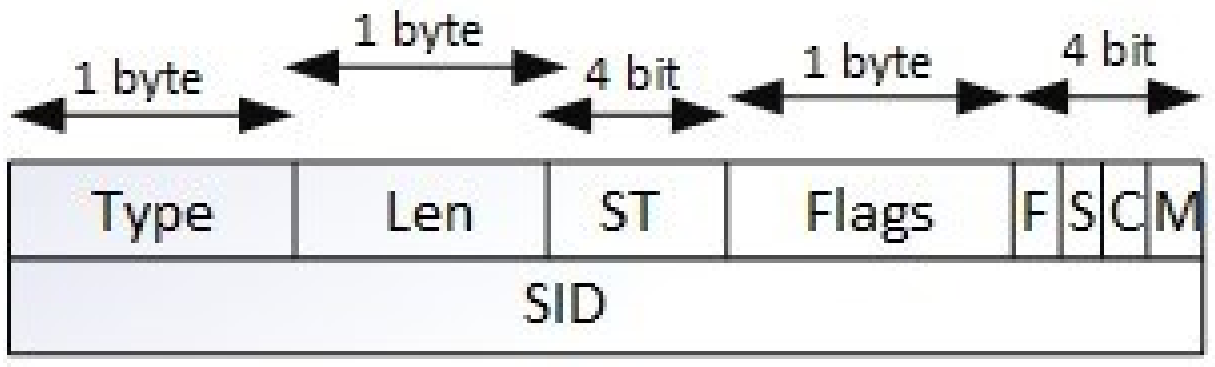

Fig. 20. SR-RRO Sub-object format.

*c) Path Computation Element Protocol (PCEP)*

Path Computation Element Communication Protocol [26] designates a PCE (Path Computation Element) that acts as a controller and calculates the path between the PCCs (Path Computation Clients). We explain the SR extension for the PCEP [27] in this subsection. For exchanging SR-capabilities, both PCC and PCE exchange messages with each other to inform that they are SR capable. This is done through including the *SR-PCE-CAPABILITY* TLV in the open messages exchange between the PCC and PCE. In PCEP, the LSP information is carried in the ERO (Explicit Route Object). Therefore, for carrying SR-TE path, a modified ERO has been defined and shown in Fig. 19 that must carry a SR-ERO sub-object. While building an MPLS stack from the ERO objects, the PCC must consider to put them in a LIFO stack. The first sub-object contains the information about the topmost label and last sub-object has the information of the bottommost label. ERO objects are sent by a PCE to a PCC. If the PCC is not segment routing capable, then it sends an error message to the PCE.

When the *L* flag is set, the PCC may overwrite one or more SID values in the SR-ERO sub-object. *ST* represents the type of information denoted by the SID tag in the sub-object. When the *ST* value is 0, then the SID must not be null and *NAI* (Node or Adjacency Identifier) must be null. If the *M* flag is set, then the SID value represents the MPLS label stack value but all other values (i.e. *TC*, *S*, and *TTL*) must then be invalid. Whereas, if the *C* flag is set along with the *M* flag, then the SID value represents the MPLS label stack and all other values (*Label*, *TC*, *S*, and *TTL*) should be populated by the PCE. When the *S* flag is set, then the SID value is null. In this case, the PCC is responsible for choosing SID value using *NAI*. If the *F* flag is set, the *NAI* value is null. The SR-ERO sub-object consist of *NAI* corresponding to the SID value. The *NAI* can be an IPv4 Node ID, an IPv6 Node ID, an IPv4 Adjacency ID or an IPv6 Adjacency ID. The value of *NAI* depends upon the value of *ST*.

The PCC records the SR-TE LSP and reports the LSP to the PCE with an RRO (Route Reply Object). The RRO must contain one or more SR-RRO sub-objects. The SR-RRO sub-object is defined in Fig. 20. The format of SR-RRO sub-object is the same as SR-ERO sub-object except for the *L* flag.